\documentclass[10,5 pt,a4paper]{article}
\usepackage{amsmath}
\usepackage{amssymb}
\usepackage{amsfonts}
\usepackage{amscd}
\usepackage{delarray}
\usepackage{graphicx}
\usepackage{color}
\usepackage{bbm}
\usepackage{authblk}
\usepackage{float}

\def\Symp#1,#2,#3,#4.{\left[\left(\begin{array}{c}#1\\#2\end{array}\right),\left(\begin{array}{c}#3\\#4\end{array}\right)\right]}
\def\Vec#1,#2.{\left(\!\begin{array}{c}#1\\#2\end{array}\!\right)}
\def\vec#1,#2.{{#1\choose{#2}}}
\newcommand{\ket} [1] {\vert #1 \rangle}
\newcommand{\bra} [1] {\langle #1 \vert}

\newcommand{\beq}{\begin{equation}}
\newcommand{\eeq}{\end{equation}}
\newcommand{\beqa}{\begin{eqnarray}}
\newcommand{\eeqa}{\end{eqnarray}}
\newcommand{\bx}{{\bf x}}
\newcommand{\by}{{\bf y}}

\newcommand{\DIS}{\displaystyle}

\definecolor{redcom}{rgb}{1,0.1,0.2}
\definecolor{querycol}{rgb}{0.2,0.2,1}

\definecolor{purplerep}{rgb}{1,0.1,1}

\definecolor{redgreen}{rgb}{0.1,0.1,1.0}

\definecolor{green}{rgb}{0.2,0.6,0.8}

\newcommand{\braket}[2]{\langle #1 | #2 \rangle}

\begin{document}
\title{de Broglie's double solution program:  90 years later}
\author[1]{Samuel Colin}
\author[2]{Thomas Durt}
\author[3]{Ralph Willox}% Force line breaks with \\
\affil[1]{Centro Brasileiro de Pesquisas F\'{\i}sicas,
Rua Dr.\ Xavier Sigaud 150,
\mbox{22290-180, Rio de Janeiro -- RJ, Brasil}
\mbox{Theiss Research, 7411 Eads Ave, La Jolla, CA 92037, USA}
}
\affil[2]{Aix Marseille Universit\'e, CNRS, Centrale Marseille, Institut  Fresnel (UMR 7249),
\mbox{13013 Marseille, France}}
\affil[3]{Graduate School of Mathematical Sciences, the University of Tokyo, 
\mbox{3-8-1 Komaba, Meguro-ku, 153-8914 Tokyo, Japan}}
\date{}
\maketitle

\vskip 1cm
\begin{abstract}
R\'ESUM\'E. {\it Depuis que les id\'ees de de Broglie furent revisit\'ees par David Bohm dans les ann\'ees '50, la grande majorit\'e des recherches men\'ees dans le domaine se sont concentr\'ees sur ce que l'on appelle aujourd'hui la dynamique de de Broglie-Bohm, tandis que le programme originel de de Broglie (dit de la double solution) \'etait graduellement oubli\'e. Il en r\'esulte que certains aspects de ce programme sont encore aujourd'hui flous et impr\'ecis. A la lumi\`ere des progr\`es r\'ealis\'es depuis la pr\'esentation de ces id\'ees par de Broglie lors de la conf\'erence Solvay de 1927, nous reconsid\'erons dans le pr\'esent article le statut du programme de la double solution. Plut\^ot qu'un fossile poussi\'ereux de l'histoire des sciences, nous estimons que ce programme constitue une tentative l\'egitime et bien fond\'ee de r\'econcilier la th\'eorie quantique avec le r\'ealisme.}

ABSTRACT. {\it Since de Broglie's pilot wave theory was revived by David Bohm in the 1950's, the overwhelming majority of researchers involved in the field have focused on what is nowadays called de Broglie-Bohm dynamics and de Broglie's original double solution program was gradually forgotten. As a result, several of the key concepts in the theory are still rather vague and ill-understood. In the light of the progress achieved over the course of the 90 years that have passed since de Broglie's presentation of his ideas at the Solvay conference of 1927, we reconsider in the present paper the status of the double solution program. More than a somewhat dusty archaeological piece of history of science, we believe it should be considered as a legitimate attempt to reconcile quantum theory with realism.}
\end{abstract}

\section{Introduction}
The Copenhagen interpretation and Everett's Many Worlds interpretation are undoubtedly very popular among the majority of quantum physicists. However, it is often conveniently forgotten that these two interpretations were severely and (most importantly) quite convincingly criticized, most notably by John Bell \cite{bell872}. Bell criticized in particular the fact that the Copenhagen interpretation presupposes the coexistence of two regimes of temporal evolution, during and in absence of a measurement.
However, the theory does not offer any objective criterion which would make it possible to infer when one of these particular regimes is valid (this is the problem of the so-called Heisenberg cut). 
Interpretations of quantum mechanics that do not recognise wave function collapse, such as the de Broglie-Bohm or many worlds interpretations, do not require Heisenberg cuts and offer some answer to the objectification problem: roughly speaking, during a measurement, one particular potentiality is actualized. Although in both theories decoherence does play a role in the answers they provide, one might still find them not fully satisfying.\footnote{In a measurement-like situation, the wave-function 
evolves into a superposition of macroscopically distinct states which will cease to interfere.
In the de Broglie-Bohm theory, the actual configuration will get trapped in one particular branch and all other branches can be neglected for all practical purposes (these are the so-called empty waves, and this aspect of the theory has been criticized many times).
In the many worlds interpretation, decoherence defines these many worlds. However, positing 
extra worlds, just to explain that each potential result gets actualized in one world, might seem excessive to some.}
It is also well-known that the decoherence program per se does not bring a satisfactory reply to the objectification problem \cite{adler-deco}.

As a result it is commonly accepted, by those who are aware of these criticisms, that the measurement problem is, for now, still a largely open problem and that no particular interpretation of quantum mechanics can be established nor privileged, solely on empirical grounds. Ultimately, the interpretation of quantum theory is still a personal choice, dictated by philosophical orientations rather than by facts. This does not mean of course that one should discard the possibility that future experimental data might, one day, allow one to discriminate between certain interpretations. This is why we think it is necessary to have as complete a picture as possible of the different plausible interpretations of quantum theory and of the experimental implications these force upon us, by which one may hope, one day, to be able to differentiate between them. This picture must include theories which, over the past decades, might not always have received their fair share of attention. One such theory is de Broglie's double solution program, which is a purely wave-monistic theory, contrary to subsequent re-formulations of the theory such as the pilot wave formulation or the Bohmian approach. As de Broglie wrote concerning the double solution program \cite{debroglie60}:

 {\it ``[A] set of two coupled solutions of the wave equation: one, the $\Psi$ wave, definite in phase, but, because of the continuous character of its amplitude, having only a statistical and subjective meaning; the other, the $u$ wave of the same phase as the $\Psi$ wave but with an amplitude having very large values around a point in space and which \ldots\ can be used to describe the particle objectively."}%{[Louis de Broglie, {\em Non-Linear Wave Mechanics -- A causal interpretation} (Elsevier, 1960)].]}\rep{Question : si on met la source de la citation `en lisible', comme on le fait ici, pourquoi on ne le fait pas de la meme facon pour la citation de Einstein, quelques lignes plus bas ? Ne suffirait-il pas de mettre (\cite{debroglie60}) juste avant, comme on fait dans le cas de Einstein ? }

It is probably not very well known however that the first attempt to solve the wave-particle duality in favour of a pure wave picture -- in fact by considering nonlinear wave equations -- can be traced back to Einstein who, in 1909, wrote the following to Lorentz \cite{letter}: 

{\it``
[T]he essential thing seems to me to be not the assumption of singular points but the assumption of field equations of a kind that permit solutions in which finite quantities of energy propagate with velocity c in a specific direction without dispersion. One would think that this goal could be achieved by a modification of Maxwell's theory. \ \ldots\ Contrary to the opinion advanced in my last publication, it seems possible to me that the differential equations that are to be sought are linear and homogeneous\ \ldots\ However, one would be forced to do the latter [i.e., introduce nonlinear or inhomogeneous equations], in my opinion, if one wished to manage without introducing singular points, which certainly would be the most satisfactory thing to do."}

This quote shows that Einstein contemplated the idea of introducing nonlinear corrections in order to obtain solitary wave solutions, and that he had some reservations  when it came to the deliberate introduction of singularities into the theory. One should keep in mind that he wrote this before the advent of general relativity, in an era still governed by linear equations.

%This quote shows that Einstein contemplated the idea of introducing nonlinear corrections in order to obtain solitary wave solutions, although he would have preferred to obtain solitary waves without the need for such corrections. \rep{Je ne sais plus qui a rajoute cette derniere remarque mais elle est clairement fausse: Einstein dit le contraire ! Il faudra repenser un peu ce paragraphe.}
%Let us keep in mind that he wrote this before the advent of general relativity, in an era governed by linear equations. 
%What needs to be retained is his first intuition about the need for nonlinear corrections in order to describe localized solutions.

Later, in the 1920's, de Broglie proposed the so-called double-solution program \cite{debroglie60} as a way of solving the puzzle of wave-particle duality, starting from the idea that particle-like properties can be explained solely in terms of waves. By the 1950's however, as a result of discussions with Vigier  (see \cite{debroglie60}, chapter VIII, section 1), de Broglie realized that his program requires a nonlinear correction to the Schr\"odinger equation.\footnote{As was noted by Vigier 
(\cite{debroglie60}, chap. VIII section 1), Einstein, in collaboration with Grommer \cite{einstein-grommer}, had a very similar objective in the framework of his theory of general relativity in the 20's: the postulate of geodetics would not be an extra hypothesis, but would be obeyed, de facto,  by peaked solutions of Einstein's nonlinear equations moving on a weakly varying metric background. 
This idea presents many deep similarities with de Broglie's guidance equation which lies at the heart of the de Broglie-Bohm hidden variable theory \cite{bohm521,bohm522}. In fact, Bohm-de Broglie trajectories are the counterpart of geodetic trajectories in Einstein's unitarian version of general relativity.
Of course, this similarity is not amazing because, as is well-known, de Broglie and Einstein never accepted the Copenhagen interpretation in the first place and favoured wave ``monism'' over wave-particle dualism in accordance with their quest for a unitary explanation of all physical phenomena.} His goal, phrased in modern language, was to explain the stability of particles in terms of {\em solitonic} properties of wave packets, for which wave-packet spreading would be counterbalanced by nonlinearities. Unfortunately, at the time, soliton theory still had to be invented. 

de Broglie also thought that the pilot-wave theory, which he first presented at the fifth Solvay conference of 1927 \cite{bacval} 
(and which was later rediscovered by Bohm in 1952 \cite{bohm521,bohm522}), was a degenerate double-solution theory in which moving soliton-like solutions have been replaced with point-particles 
(see \cite{Fargue} for a review and \cite{fargue-proc}). 
More explicitly, the pilot-wave theory, in its non-relativistic version, says that a single particle is not only described by a wave-function $\psi(t,{\bf x})$ but also by a position ${\bf x}(t)$. 
The wave-function always evolves according to the Schr\"odinger equation, whereas the velocity of the particle is given by the gradient of the phase of the wave-function 
(evaluated at the actual position of the particle), but the exact mechanism through which this coupling occurs is not made precise (contrary to the double solution program in which the precise description of the coupling of the $u$ and $\psi$ waves is one of the main stumbling blocks to the development of such a theory). An ensemble of particles, on the other hand, is described by the wave-function $\psi(t,{\bf x})$ and by a distribution 
of particle positions, denoted by $\rho(t,{\bf x})$.  A priori $\rho(t,{\bf x})$ is not related to the standard quantum distribution $|\psi(t,{\bf x})|^2$, 
but the pilot-wave theory reproduces the predictions of standard quantum mechanics only if $\rho(t,{\bf x})=|\psi(t,{\bf x})|^2$. 
If this last condition holds at some initial time, it will hold for any later time, which is why such a distribution is referred to as a quantum equilibrium distribution.
The theory can be generalized to many-particle systems \cite{debroglie27,bohm521} and it can also be extended in order to reproduce the predictions of standard quantum field theory \cite{bohm522}
(in that case, fermionic particle positions and bosonic field configurations, together with wave functionals, provide a complete description of the universe).

Although in 1927 nobody fully understood the profoundness of the conceptual problems linked to the interpretation of quantum theory \cite{bacval}, de Broglie's attempt \cite{debroglie27} to derive a many particle guidance equation already contained germs of what would later be called the {\em nonlocality} issue. In \cite{debroglie27}, de Broglie struggled with the necessity to formulate the guidance equation in configuration space, rather than in ``real'' 3-dimensional space. It was only in 1935, with the infamous EPR and Schr\"odinger cat paradoxes that the issues of nonlocality and the measurement problem
were explicitly identified.\footnote{It should be pointed out that, in 1926, Born  \cite{born} also struggled with entanglement, when he described a typical scattering experiment between a projectile and a target. Nowadays this is well-known: after $A$ is scattered by $B$, the full system $A$-$B$ is highly entangled because of conservation of momentum: the recoil of $B$ must compensate the change of momentum of $A$ \cite{zeit}. At the time however, Born settled this problem by conditioning the $A$-states on the $B$-states, an idea that reappears much later in the framework of the modal interpretation. It is also worth remembering that Bohm's personal contribution, 25 years later, to the pilot wave theory was in fact double: he showed explicitly how in its non-relativistic formulation, and in an EPR like approach, the guidance equation is nonlocal. Moreover, he also understood that what is nowadays called decoherence, actually provides a tool to solve the measurement problem, when added to the pilot wave approach.}
 
Since de Broglie's pilot wave theory was revived by David Bohm in the 1950's, the overwhelming majority of researchers involved in the field have focused on what is nowadays called de Broglie-Bohm dynamics and de Broglie's original double solution program was gradually forgotten. 
As a result, several of the key concepts in the theory are still rather vague and ill-understood. 
For example, de Broglie himself, during his life time,  at times thought of particles as singular solutions to the Schr\"odinger equation (see e.g. Ref. \cite{vigier} and \cite{fargue-proc}), but sometimes also as peaked, but finite, (hump or lump like) waves. 
de Broglie's vacillation on this topic has a modern equivalent in the confusion that surrounds so-called rogue waves in present-day nonlinear science: are these indeed best described by peaked but finite, particular solutions to certain nonlinear model equations, or might there be cases where they are in fact mistaken for blow up-type solutions to such equations? These problems are the topic of section \ref{RW-rogue}, where different types of rogue and solitary waves and their relevance to the double solution program will be discussed.

Over the past 50 years, enormous progress has been made in the study of solitonic solutions to nonlinear PDEs {(cf. section \ref{RW})}, as a result of which various nonlinear generalisations of the Schr\"odinger equation have been  studied. Fargue \cite{Fargue} has shown that in the {\em free} case there exist several nonlinear PDEs that admit finite solitonic solutions (localised humps) propagating in accordance with the de Broglie-Bohm guidance equation. These equations are labeled `free' in the sense that, apart from a self-focusing nonlinearity, no external potential acts on the particle. Moreover, all of these equations are Galilei invariant, which provides an alternative explanation (see \cite{durt-single}, section 2, and \cite{durt-proc}) of why de Broglie's guidance relation is fulfilled in the free case. To our knowledge however, the original double solution program has never been fully realised, despite the sporadic proposals of nonlinear generalisations  of the Schr\"odinger equation that have been made over time. In section \ref{NLmodif}, we give a partial historical survey of these attempts. 

In parallel to these {largely} inconclusive attempts, several `no-go' theorems also inhibited research in the direction outlined by de Broglie in 1927. In the present paper we shall discuss two of these: Derrick's no go theorem \cite{Derrick} related to the stability of de Broglie like solitary waves (in section \ref{RW-nogo}) and Gisin's theorem linking nonlinearity and nonlocality (section \ref{nogo2}). As we have shown in a previous paper \cite{CDW}, Derrick's no-go theorem suffers from  a basic flaw, which is that it implicitly assumes that the evolution equation is not norm-preserving (a sketch of the reasoning can be found in appendix C). Moreover, as we show in section \ref{nogoBorn}, Gisin's theorem can be circumvented provided the spatial distribution of de Broglie solitons obeys the Born rule. The dynamics however remains nonlocal, just as in the case of certain classical nonlinear PDEs, which we shall discuss in section \ref{RW-unstable} in relation to the role played by conservation laws and resonant solutions for such equations.

In the light of these insights and of the progress achieved over the course of the 90 years that have passed since de Broglie's original ideas, we hope we are now finally able to reconsider the status of the double solution program. More than a somewhat dusty archaeological piece of history of science, we believe it should be considered as a legitimate attempt to reconcile quantum theory with realism\footnote{There have of course been other attempts in this direction, that do not fit nicely with the wave monism of de Broglie. Here we have in mind Bohm's {theory in which particles are represented in terms of material points}, or spontaneous localisation models, where wave particle duality is implicitly assumed to be present from the beginning. One may wonder whether de Broglie's double solution program might not be the last hope for a theory reconciling quantum theory and realism, that respects wave monism.}. An essential ingredient of this program, in our eyes, is nonlinearity. In this sense it differs radically from the text-book formulations of quantum theory. In accordance with Bell's analysis and Gisin's no-go theorem, this nonlinearity is accompanied by nonlocality and the violation of Lorentz covariance, which are also important departures  from the mainstream formulation of modern physics. As we shall discuss, nonlocality (in the sense of no-signaling (section \ref{nogomust})) seems to be the price to pay if one wishes to adopt a realistic interpretation of quantum theory  (section \ref{nogoreal}), in full agreement with Bell's analysis which says that local realism is excluded by the violation of Bell's inequalities, while nonlocal realism most definitely is not. 

On the other hand -- leaving aside questions of interpretation for a moment -- it could in fact very well be that the ``missing link'' between quantum theory and general relativity (which is fundamentally a nonlinear theory) is in fact nonlinearity, as was already suspected by J-P Vigier several decades ago (from de Broglie's recollection in \cite{debroglie60}). If this is indeed the case, it is imperative to pursue de Broglie's program in depth in the context of what are nowadays called `semi-classical' gravitation theories (an overview can be found in appendix B). 
Such a study might contribute to the dissolution of the schism between quantum theory and gravitation theories, which constitutes the most pressing theoretical challenge in today's physics. 

 It is worth noting that, quite recently, de Broglie's point of view has been revived, be it indirectly, by experimental observations in hydrodynamics, which show that certain macroscopic objects, so-called walkers (bouncing oil droplets), exhibit many of the features of the de Broglie-Bohm (dB-B) dynamics \cite{couder1,couder2,couder3,bush}. These unexpected developments not only show that de Broglie's ideas encompass a large class of systems, but they might in the future also allow us to build a bridge between quantum and classical mechanics, where ingredients such as nonlinearity, solitary waves and wave monism play a prominent role (see also \cite{durt-epl} and \cite{borghesi-proc}).

\section{Historical developments surrounding the double solution program}\label{NLmodif}
In this section, we provide an overview of some important ideas relevant to the double solution program. These include developments in nonlinear 
quantum mechanics, but also in general relativity and 
in interpretations of standard (linear) quantum mechanics such as the pilot-wave theory.
The presentation of these contributions follows their timeline and is definitely not exhaustive.
\subsection{de Broglie 1923--1927}
The idea of the Broglie, back in 1923, was that propagation of a wave should be associated to any corpuscular motion. Considering a corpuscle or particle as a tiny clock, de Broglie realized that, in the simplest case of free motion, the only way for the particle and the monochromatic wave to remain in phase is 
if the energy (momentum) of the particle is equal to the wave frequency (number) multiplied by a universal constant (found to be the Planck constant).
These now famous de Broglie relations were first verified experimentally by Davisson and Germer in 1927, for electrons. To de Broglie's mind, only the phase of the wave (whose amplitude is the same everywhere in this simple case) had physical significance, which led him to call it the {\em phase wave}. 
Subsequently, Schr\"odinger found a general equation of motion for the wave (called alternatively the $\psi$-wave or quantum wave-function). However, later on, 
due to the development of the probabilistic Copenhagen interpretation by Bohr, Heisenberg and Born, the particle itself disappeared entirely along the way.\footnote{It is actually stronger than that: the realistic character of the interpretation disappeared entirely.
Wave-monism does not necessarily preclude a realistic interpretation (Schr\"odinger's idea).
However, wave-monism {\em plus} linearity does, in agreement with Born's analysis \cite{born}.}
This was highly dissatisfying for de Broglie and in his effort to reinstate the role of the corpuscle in the theory, he started to develop his double solution program (1925-1927), the (first) core idea of which is that there should be two synchronous, coupled, solutions of the wave equation: 
\begin{itemize}
\item a $\psi$-wave, the phase of which has physical significance but the amplitude of which does not, 
\item and a $u$-wave, which is  a solution describing a moving singularity, the singularity corresponding to the particle. 
A trivial example would be $u(t,{\bf x})=\frac{1}{|{\bf x}-{\bf x}(t)|}e^{i S(t,{\bf x})/\hbar}$.

The phase of the $u$-wave is required to be equal to that of the $\psi$-wave and the $u$-wave itself is supposed to provide a description of the physical reality.
\end{itemize}
The second main idea of the program is that the singularity described by the $u$-wave should follow the flow lines of the wave equation. de Broglie in fact also wanted to get rid of configuration space, but we shall come back to this idea later in this section. 
For the time being, it suffices to say that the difficulties de Broglie \cite{debroglie27} was confronted with arose in his analysis of two interacting spinless particles.\footnote{As mentioned in the introduction, the difficulties met in the two-particle case are precursors 
of the nonlocality problem, recognized some years later in the EPR paper, and fully recognized by John Bell only several decades later.}
According to the double-solution program, there should then be two $u$-waves, one for each particle: $u_1(t,{\bf x})$ and $u_2(t,{\bf x})$. 
In order to obtain the equation satisfied by both $u_1$ and $u_2$, de Broglie assumed that each $u$-wave would feel the potential generated by the other. 
More precisely, in the presence of an external potential $V=V({\bf x})$, the Klein-Gordon equation becomes
\begin{equation}
\Delta u-\frac{1}{c^2}\frac{\partial^2 u}{\partial t^2}
+\frac{2i}{\hbar c^2}V\frac{\partial u}{\partial t}-\frac{1}{\hbar^2}(m^2c^2-\frac{V^2}{c^2})u=0,
\end{equation}
(the equation obtained from the free K-G equation by replacing the energy 
operator by $E+V$).

de Broglie therefore assumed that the system of coupled equations was the following one
\begin{align}
\Delta u_1-\frac{1}{c^2}\frac{\partial^2 u_1}{\partial t^2}
+\frac{2i}{\hbar c^2}V_1\frac{\partial u_1}{\partial t}-\frac{1}{\hbar^2}(m_1^2c^2-\frac{V_1^2}{c^2})u_1=0,\label{eq_u1}\\
\Delta u_2-\frac{1}{c^2}\frac{\partial^2 u_2}{\partial t^2}
+\frac{2i}{\hbar c^2}V_2\frac{\partial u_2}{\partial t}-\frac{1}{\hbar^2}(m_2^2c^2-\frac{V_2^2}{c^2})u_2=0,\label{eq_u2}
\end{align}
where $V_1=V(|\bf{x}-\bf{x_2}(t)|)$ and $V_2=V(|\bf{x}-\bf{x_1}(t)|)$, ${\bf x_1}(t)$ and ${\bf x_2}(t)$ being the positions of the singularities and $V$ being a classical potential (a Coulomb or Yukawa potential for example). Here $u_1$ and $u_2$ are now assumed to be of the form $u_1(t,{\bf x},{\bf x_2}(t))$ and $u_2(t,{\bf x},{\bf x_1}(t))$.

However, de Broglie did not attempt to solve this complex system. 
Instead, as a preliminary theory, de Broglie replaced each $u$-wave by its position. 
On the basis of the existence, in classical mechanics, 
of a function $\phi({\bf x_1},{\bf x_2})$ 
for which the velocities are proportional to the gradients of $\phi$, 
he assumed that this is also the case in quantum mechanics.
In non-relativistic quantum mechanics, the last piece of the theory then comes from the identification of this function $\phi$ with the phase of a function $\Psi$ which satisfies a 2-body Schr\"odinger equation. 
Doing so, he obtained what is now referred to as the de Broglie-Bohm pilot-wave theory, a theory which he saw as a mere approximation of the double-solution, but which he nevertheless presented at the fifth Solvay congress (a detailed analysis of \cite{debroglie27} can be found in \cite{bacval}). 
At the Solvay conference, the theory faced serious objections and de Broglie soon rallied to the Copenhagen camp. (de Broglie writes about this in the preface of his book \cite{debroglie60}, or in \cite{debroglie59} for example.)
\subsection{Darmois - Einstein 1927}
In 1927, de Broglie still lacked a crucial ingredient for his program: the fact that $\psi$ should obey a nonlinear wave-equation. In hindsight, this problem should of course be viewed in parallel to the contemporary developments that took place in general relativity.

Around the same time, Einstein and Grommer, and Darmois independently 
(see \cite{debroglie59}, page 971, \cite{infeld-schild} and references therein), 
tackled the following problem. 
In general relativity, one has the nonlinear Einstein field equations and it is further assumed that a corpuscular test particle follows a geodesic. 
The question is whether the geodesic equation of motion can be derived from the Einstein field equations alone, if we interpret the corpuscular test particle as some kind of singularity of the field. 
In some cases it is indeed possible to show that, starting 
from a solution of Einstein's field equations, 
it is possible to add to this solution a function describing a moving singularity and still obtain a solution, provided that the singularity follows the geodesic equation of motion. 
In other words, the field equations impose a particular guidance equation on the particle (seen as a singularity in the field).
\subsection{Bohm 1952}
The starting point of Bohm \cite{bohm521} was to use the polar decomposition 
\beq
\Psi(t,{\bf x_1},\ldots,{\bf x_N})=R(t,{\bf x_1},\ldots,{\bf x_N})e^{\frac{i}{\hbar}
S(t,{\bf x_1},\ldots,{\bf x_N})}
\eeq
in the Schr\"odinger equation
\beq
i\hbar\partial_t\Psi=
-\sum_{n=1}^{n=N}\frac{\hbar^2}{2 m_n}\Delta_n\Psi+V(t,{\bf x_1},\ldots,{\bf x_N})\Psi,
\eeq
where $m_n$ is the mass of the $n$-th particle and $V$ a classical potential.
This leads to a continuity equation for $R^2$ and a Hamilton-Jacobi equation, if we admit 
that there is an extra potential, the quantum potential
\beq\label{qpot}
Q(t,{\bf x_1},\ldots,{\bf x_N})=-\frac{\hbar^2}{2 R}\DIS\sum_{n=1}^{n=N}\frac{\Delta_n R}{m_n},
\eeq
and if we assume that the velocity of each particle is given by the gradient of the phase, with respect to the corresponding coordinate, divided by the mass of the particle. 

The theory can also be formulated in a quasi-Newtonian fashion, if the acceleration 
of each particle is given by
\beq\label{acc-bohm}
{\bf a}_{n}(t)=-\frac{1}{m_n}{\bf\nabla_n}(V+Q)\bigg|_{{\bf x_1}={\bf x_1(t)},\ldots,{\bf x_N}={\bf x_N(t)}}.
\eeq 

This dynamics reproduces the predictions of the standard interpretation of quantum mechanics,
provided that the positions of the particles are initially distributed according to Born's law 
over an ensemble, and provided that the velocities are initially given by
\beq
{\bf v_n}(t_i)=
\frac{1}{m_n}{\bf\nabla_n} S(t_i,{\bf x_1},\ldots,{\bf x_N})\bigg|_{{\bf x_1}={\bf x_1}(t_i),\ldots,{\bf x_N}={\bf x_N}(t_i)}.\label{v-phase}
\eeq
Note that \eqref{v-phase} does not have quite the same status as \eqref{acc-bohm}: \eqref{v-phase} is an initial condition which can in principle be relaxed (see \cite{cova}) whereas \eqref{acc-bohm} is a dynamical law.

Another major contribution of Bohm was the detailed analysis of a measurement situation \cite{bohm522}, in order to explain the apparent reduction of the wave-function. This analysis can be sketched as follows.

Let us denote by $x$ (resp. $y$) the collection of positions belonging to a system (resp. apparatus).
We want to measure some observable $\widehat{A}$ of the system thanks to the apparatus.
The initial wave-function $\Psi(t_i,x,y)$ is a product of the wave-function of the system with that of the apparatus
\beq
\Psi(t_i,x,y)=\psi_{sys}(x)\psi_{app}(y)=(\sum_n c_n a_n(x))\psi_{app},
\eeq
where the $a_n(x)$ are the eigenstates of $\widehat{A}$.
A `measurement' amounts to turning on some interaction Hamiltonian 
between the system and the apparatus, which will 
correlate the eigenstates $a_n(x)$ of $\widehat{A}$ to eigenstates $\phi_n(y)$ of the apparatus (these $\phi_n(y)$ are  macroscopically distinct states, like pointer up or pointer down states). Therefore, 
at the end of the measurement, the wave-function is
\beq\label{branches}
\Psi(t_f,x,y)=\sum_n c_n a_n(x)\phi_{n}(y)=\sum_n c_n\Phi_n(x,y)~,
\eeq
where the states $\Phi_n$ are the different branches of the wave-function, 
each branch corresponding to a potential result of the measurement. 
These branches have almost no overlap in the position basis $(x,y)$. 
During the measurement, the actual configuration 
$(x(t),y(t))$ finally becomes trapped in one branch, say $\Phi_{n_0}$, because the different branches are separated by regions where $|\Psi|^2\approx 0$ in which the configuration cannot move into.
Inside the branch $n_0$, the actual configuration will only feel the influence of $\Phi_{n_0}(x,y)$ 
as far as the guidance equation (\ref{acc-bohm}) is concerned, which explains why 
we can discard the other branches (which are therefore often referred to as empty waves).
It is also virtually impossible for the branches to overlap subsequently, 
because of the interaction of the apparatus with the environment (an overlap would also require an overlap in $z$ -- where $z$ collectively denotes the positions of the particles belonging to the environment -- which is practically impossible). 
This is how the theory answers the objectification problem mentioned in the introduction. 

In the same article \cite{bohm522}, Bohm even proposed an extension to relativistic bosonic fields. 
\subsection{de Broglie after 1952: some remarks}
The work of Bohm caused a renewal of de Broglie's interest in his abandoned double-solution program. 
In particular, de Broglie introduced a new hypothesis, which he considered as indispensable, which is the nonlinearity of the equation of motion, i.e.: there should be a nonlinear correction  $V_{NL}$ to the Schr\"odinger equation
\begin{equation}
i\hbar\partial_t\psi=-\frac{\hbar^2}{2m}\Delta\psi+V_{L}\psi+V_{NL}\psi,
\end{equation}
which should only be important for peaked solutions (in order to preserve the success of the standard interpretation). (For the time being we only consider the case of a single particle.)

$V_{NL}$ will contain a parameter, let us call it $g$, 
which measures the strength of the coupling to $\psi$ (for example as in $V_{NL}\psi=-g|\psi|^2\psi$).
There are, in fact, strong experimental bounds on the value of $g$ (see for example \cite{bbmy,smolin,bb}).

We shall now make a few crucial remarks concerning the double solution program 
in the light of this new hypothesis of nonlinearity.
\subsubsection{Remark 1: the connection to Einstein's ideas}
Let us assume that we have a $\psi$-wave solution to the nonlinear equation, which is approximately equal to a standard quantum solution, i.e. to a solution to the linear Schr\"odinger equation. 
Let us now consider a $u$-wave which is possibly singular around a moving point.
It is assumed that the $u$-wave has the same phase as the $\psi$-wave: 
$\psi(t,{\bf x})=R(t,{\bf x}) e^{i  S(t,{\bf x})/\hbar}$. One possible form for $u$ is then 
$r({\bf x}-{\bf x}(t))e^{i \hbar^{-1} S(t,{\bf x})}$ (where the function $r$ can be singular at the origin).
Alone, just by itself, $u$ is not a solution.
The hope of de Broglie, however, was that the sum of $u$ and $\psi$ 
would be a solution, provided that ${\bf x}(t)$ moves along the flow lines, that is provided that
\beq
{\bf v}(t)=\frac{1}{m}{\nabla S(t,{\bf x})}\bigg{|}_{{\bf x}={\bf x}(t)}.
\eeq
The guidance equation would therefore arise from the nonlinear wave equation in the same way as the geodesic equation of motion arises from the Einstein field equations in general relativity.

This amounts to showing that the following equation is satisfied:
\begin{eqnarray}
i\partial_t(r({\bf x}-{\bf x}(t))e^{iS})=(-\frac{1}{2m}\Delta+V_{L})r e^{iS}\nonumber\\
+V_{NL}[(R+r)e^{iS}](R+r)e^{iS}-V_{NL}[Re^{iS}]Re^{iS},
\end{eqnarray}
(where $\hbar=1$) provided that $m\frac{d{\bf x}(t)}{dt}=\nabla S(t,{\bf x})$.

de Broglie never gave an explicit form of the nonlinear correction $V_{NL}$ to the Schr\"odinger equation (for an explicit proposal, see \cite{vigier-pla,Fargue} and appendices B and C).
However, once we specify a nonlinear term, the above question can in principle be investigated.

\subsubsection{Remark 2: from configuration space to physical space}
In the standard interpretation of quantum mechanics, the wave-function $\Psi(t,{\bf x_1},\ldots,{\bf x_N})=R e^{i\hbar^{-1}S}$ describing a system of $N$ particles is defined on a configuration space of dimension $3 N$. 
Statistical predictions can be obtained from this function $\Psi$. 
In particular, for a system of 2 particles the probability density to find particle 1 at position ${\bf x_1}$ and particle 2 at position ${\bf x_2}$, is given by $R^2(t,{\bf x_1},{\bf x_2})$, and we have the continuity equation
\begin{equation}\label{conteq1}
\frac{\partial R^2}{\partial t}+\nabla_1\cdot(R^2 \frac{{\bf \nabla_1} S)}{m_1})
+\nabla_2\cdot(R^2 \frac{{\bf \nabla_2} S)}{m_2})=0.
\end{equation}
In the double solution model, however, everything should be expressed in physical space.
The question is then how to reproduce the predictions of the standard interpretation.
In the following we present some insights into this question 
from de Broglie and his collaborator Andrade e Silva. 
These are supposed to be valid for the double solution program and the pilot-wave theory.

For simplicity, let us consider the two-particle case. 
Instead of having the wave-function $\Psi(t,{\bf x_1},{\bf x_2})$, we have $2$ functions, 
$\nu_1$ and $\nu_2$ (one for each particle), defined on the physical space. 
It is assumed that these functions are of the form
\beq
\nu_1(t,{\bf x},{\bf x_2}(t)),\quad \nu_2(t,{\bf x_1}(t),{\bf x}),
\eeq
where ${\bf x_1}(t)$ and ${\bf x_2}(t)$ are the positions of the moving singularities (or particles). 
Let us call $r_1$ and $r_2$ the respective amplitudes, where 
$r_1=r_1(t,{\bf x},{\bf x_2}(t))$ and $r_2=r_2(t,{\bf x_1}(t),{\bf x})$.
Each $\nu_n$ satisfies a wave-equation, which is not yet specified however: 
all that is assumed is that there are continuity equations 
for $r^2_1$ and $r^2_2$:
\begin{align}\label{conteq2}
\frac{\partial r^2_1}{\partial t}+{\bf \nabla}\cdot(r^2_1 {\bf v_1})=0,\qquad\qquad
\frac{\partial r^2_2}{\partial t}+{\bf \nabla}\cdot(r^2_2 {\bf v_2})=0.
\end{align}
In their attempt to derive configuration-space equations from those in physical space, 
de Broglie and Andrade e Silva define $R(t,{\bf x_1},{\bf x_2})=\tilde{r}_1(t,{\bf x_1},{\bf x_2})\tilde{r}_2(t,{\bf x_1},{\bf x_2})$, 
where $\tilde{r}_1(t,{\bf x_1},{\bf x_2})$ is the expression obtained from $r_1(t,{\bf x},{\bf x_2}(t))$ by replacing ${\bf x}$ by ${\bf x_1}$ and ${\bf x_2}(t)$ by ${\bf x_2}$ 
(similarly for $\tilde{r}_2(t,{\bf x_1},{\bf x_2})$). 
They then go on to show \cite{acasci1} that $R^2$ satisfies the above continuity equation (\ref{conteq1}), 
because of (\ref{conteq2}).

So far, it has only been assumed that $\nu_1$ and $\nu_2$ have associated continuity equations, but their respective equations of motion have not been specified yet. 
de Broglie and Andrade e Silva first assumed that the equations were of the form
\begin{align}
&i\hbar\partial_t\nu_1=-\frac{\hbar^2}{2m_1}\Delta\nu_1+V({\bf x}-{\bf x_2}(t))\nu_1,&\\ 
&i\hbar\partial_t\nu_2=-\frac{\hbar^2}{2m_2}\Delta\nu_2+V({\bf x_1}(t)-{\bf x})\nu_2.&
\end{align}
But these equations are too simple and they ran into difficulties in their attempt to reproduce the standard predictions. They therefore assumed that 
\begin{align}
&i\hbar\partial_t\nu_1=-\frac{\hbar^2}{2m_1}\Delta\nu_1+V({\bf x}-{\bf x_2}(t))\nu_1
+V'_1({\bf x}-{\bf x_2}(t))\nu_1,& \label{nonlinqp1}\\
&i\hbar\partial_t\nu_2=-\frac{\hbar^2}{2m_2}\Delta\nu_2+V({\bf x_1}(t)-{\bf x})\nu_2
+V'_2({\bf x_1}(t)-{\bf x})\nu_2,\label{nonlinqp2}&
\end{align}
where $V'_1$ and $V'_2$ are potentials of quantum interaction. In a series of notes \cite{acasci2,acasci3,acasci4}, Andrade e Silva considers cases of increasing complexity (from 2 to 3 particles).
What is clear from these notes is that the new terms $V'$ become more and more complex. 
More importantly, however, is that the equations for the $\nu_n$ are nonlinear ones, as 
can be inferred from (\ref{nonlinqp1}) and (\ref{nonlinqp2}) because the quantum potential $V'$, like (\ref{qpot}), depends on the amplitudes of the waves.
It also seems that the case of indistinguishable particles was never treated.

This approach bears some resemblance to the notion of conditional wave-functions and 
to the attempt of Norsen to express the de Broglie-Bohm pilot-wave theory in physical space \cite{norsen}.
However this approach is, in fact, significantly different in the sense that Norsen proceeds the other way around: he derives the equations valid in physical space from 
the one(s) that one has in configuration space. 
If one has a wave-function $\Psi(t,{\bf x_1},{\bf x_2})$ and two positions ${\bf x_1}(t)$ and 
${\bf x_2}(t)$, the starting point of Norsen's approach is to define 
two conditional wave-functions
\begin{align}
&\psi_1(t,{\bf x})=\Psi(t,{\bf x_1},{\bf x_2})\bigg{|}_{{\bf x_1}={\bf x},{\bf x_2}={\bf x_2}(t)},&\\
&\psi_2(t,{\bf x})=\Psi(t,{\bf x_1},{\bf x_2})\bigg{|}_{{\bf x_2}={\bf x},{\bf x_1}={\bf x_1}(t)}.&
\end{align}
Since the equations of motion for $\Psi$ and for the positions ${\bf x_1}(t)$ and 
${\bf x_2}(t)$ are known, the equation of motion for the conditional wave functions can be obtained. If one insists on writing the equation in terms of local beables, it becomes a complicated equation with many new fields whose role is to enforce entanglement. In particular, it is non-unitary.

The lesson to be drawn from this remark is that the wave equations in physical space become 
highly complicated and in particular they become nonlinear 
(even if there is no nonlinear correction to the Schr\"odinger equation on configuration space).
\subsubsection{Remark 3: the origin of the Born rule}
If the particle positions are initially distributed according to $|\psi(t_i,{\bf x})|^2$ over an ensemble, 
they will be distributed according to $|\psi(t_f,{\bf x})|^2$ for any later time. 
This is a consequence of both the continuity equation 
and of the fact that the velocity of the particle is equal to the quantum mechanical 
current divided by $|\psi|^2$.
A concern of de Broglie was the justification of $|\psi|^2$ as a probability density 
(echoing an early  objection of Pauli to the pilot-wave theory). 
Two different answers have been provided to this question (see \cite{drezet-proc}).

One is due to Bohm and Vigier \cite{bohm-vigier} who assumed the existence of a sub-quantum fluid  and showed that it can drive an out-of-equilibrium distribution 
($\rho\neq|\psi|^2$) to equilibrium ($\rho=|\psi|^2$) thanks to stochastic fluctuations.

Another answer is due to Antony Valentini \cite{valentini-phd} and it does not suppose any modification of the quantum theory. 
Valentini and Westman \cite{valentini042} have simulated the evolution of a non-equilibrium distribution for a particle whose wave-function is a superposition of the first 16 modes of energy 
(over a domain that is a two-dimensional square box). 
They have shown that the distribution quickly relaxes to equilibrium on a coarse-grained level. 
This relaxation to quantum equilibrium was expected from earlier theoretical results. 
In \cite{valentini92}, the coarse-grained H-function is defined:
\beq
\bar{H}(t)=\int d q_1\ldots dq_{3N}\bar{\rho}\ln{(\frac{\bar{\rho}}{\overline{|\psi|^2}})}
\eeq  
where the bar over a quantity denotes a coarse-graining. 
It is shown that $\frac{d\bar{H}(t)}{dt}\leq 0$. Since the minimal value of that function (zero) can only be reached if $\bar{\rho}=\overline{|\psi|^2}$, it indicates a tendency to go to equilibrium.
Since then, many more numerical simulations have been performed in order to study various aspects of the relaxation process: 
how modifications of the pilot-wave theory affect the relaxation \cite{cost10}, 
how  the relaxation time scales with the number of modes \cite{toruva}, 
whether relaxation to quantum equilibrium occurs for fermions \cite{colin2012}, 
how it behaves for longer times \cite{abcova}, as well as the identification of systems with residual non-equilibrium \cite{efthymiopoulos3,abcova}.
The origin of the relaxation process has also been better understood since then 
(see \cite{efthymiopoulos,efthymiopoulos2} and \cite{efthymiopoulos-proc}).
\subsubsection{Remark 4: the normalization of the wave-function\label{secwein}}
Independently from de Broglie's double solution program, several attempts have been made to add 
a nonlinearity to the Schr\"odinger equation (see for example \cite{bbmy,goldin,miret} and the first two appendices to the present paper).
One relevant example for this remark is a formalism due to Steven Weinberg.

In 1989, Weinberg enlarged the formalism of standard quantum mechanics in order to allow the possible existence of nonlinear corrections to the Schr\"odinger equation \cite{weinberg1}. 
His goal was to devise experimental tests for the existence of such corrections \cite{weinberg2}.
However his formalism is very restrictive, because it does not aim to explain the reduction of the wave-function and there is still a collapse postulate in the formalism. 
As such, Weinberg can only consider nonlinear corrections for which $Z\psi$ is a solution if $\psi$ 
is a solution (where $Z$ is an arbitrary complex number). Such nonlinear corrections are called 
homogeneous and they allow for a rescaling after a measurement has occurred. 
Weinberg puts it very clearly that the nonlinear Schr\"odinger (NLS) equation (which is similar to the Schr\"odinger-Newton equation discussed in appendix B and is also introduced in appendix C)
\beq
i\partial_t\psi=-\frac{1}{2m}\Delta\psi+\epsilon|\psi|^2\psi
\eeq
does not belong to his formalism.

However, if positions exist together with the wave-function, as in the de Broglie-Bohm pilot-wave theory, 
such a restriction can be overcome altogether. To make this claim clear, let us sketch such a theory.

A system would be described by a collection of particles positions $x(t)$ 
and by a wave-function $\psi(t,x)$ obeying a nonlinear wave equation.
An ensemble, on the other hand, is described by a wave-function $\psi(t,x)$ and by a distribution of particle positions $\rho(t,x)$. 
For many nonlinear equations (such as the NLS equation or the Schr\"odinger-Newton equation), the continuity equation is identical to that of the linear Schr\"odinger equation.
Therefore, if the particles move according to the standard guidance equation and 
if we start from a distribution such that 
$\rho(t_i,x)=K|\psi(t_i,x)|^2$, it will be of the form 
$\rho(t_f,x)=K|\psi(t_f,x)|^2$ for any later time $t_f$ (since in this case it is not possible to rescale the solution and obtain another one, a constant factor $K$ needs to be introduced to ensure that $\rho$ is normed to $1$).

In a measurement-like situation, 
we could end up with a superposition of branches $\displaystyle\sum_n c_n\Phi_n(x,y)$ (similarly to (\ref{branches})).
Another possibility would be for the wave-function to collapse naturally to one state $\Phi_{n_0}(x,y)$ 
(for example, in the case of a gravitational collapse of the wave-function).
In any case, the actual configuration will get trapped inside one branch and will only 
feel the influence of this branch as far as the guidance equation goes. 
If we repeat the experiment many times, 
what is the probability to find the result $n_0$? 
Over many experiments the position configurations will be distributed according to  $\rho(t_f,x,y)=K|\Psi(t_f,x,y)|^2$ and the probability to see the result $n_0$ is equal to the probability of the configuration 
being in the support of $\Phi_{n_0}(x,y)$, which is $|c_{n_0}|^2$.
 
Notice that the absence of rescaling does not cause any problem for a pilot-wave theory with nonlinear evolution of the wave-function, for the simple reason that $|\psi|^2$ is not a probability density and does not need to be normed to 1 ($\rho$ is the probability density). The difference between 
$|\psi|^2$ and $\rho$ was made very clear in the work of Valentini on the possible existence of quantum non-equilibrium. Regarding the latter, 
if we depart from $\rho(t_i,{\bf x})=K|\psi(t_i,{\bf x})|^2$, it is worth investigating
whether there is a relaxation to $\overline{\rho(t_f,{\bf x})}=K\overline{|\psi(t_f,{\bf x})|^2}$ (where,as before, the tilde denotes a coarse-graining), despite the fact that $\psi$ undergoes a nonlinear evolution.

Such reasoning can in fact be carried over to the double-solution program. 
In conclusion, there is in fact no need for $\psi$ to be normalized, whether $\psi$ obeys a linear or a nonlinear equation in the double-solution (be it the watered-down pilot-wave theory or not).

\section{Nonlinearity and nonlocality in a classical context} \label{RW} Dispersion is a well-known property of solutions of linear wave equations. It is also the main reason why  the stability of particles, if interpreted as waves, cannot be explained within in the framework of the linear wave mechanics of Schr\"odinger or de Broglie. This apparent failure of a pure (in this case linear) wave picture then leaves the door wide open to dualist interpretations {\it \` a la} Copenhagen. In the framework of nonlinear partial differential equations however the situation is quite different, and many examples are known of nonlinear wave equations that admit solitary waves or even solitons as exact solutions. As a result, there exists a long tradition in mathematical physics and particle physics to consider solitons or in a broad sense, soliton-like, localized quasi-stable solutions to certain field equations, as quasi-particles or effective particles (see \cite{Manton} for a thoughtful review and e.g. \cite{Cern-on} for a recent extension of the soliton-particle paradigm to supersymmetric gauge theories). This approach is motivated not only by the fact that solitons are such that their spreading due to dispersion is exactly compensated by nonlinear effects, but also by the fact that solitons can undergo collisions, cross each other etc, without losing their `individuality': after a while the number of ``bumps'' as well as the heights and speeds of these bumps in a multi-soliton solution are the same as before the collision. This seems to suggest that they would be ideal candidates for realizing de Broglie's program.
However, the remarkable stability properties with respect to mutual interactions exhibited by genuine solitons, as opposed to mere {\em traveling} or {\em solitary waves} (i.e., localized hump-like waves that evolve without changing shape but which lack the stability of the solitons w.r.t. to mutual interactions), those properties only occur in integrable models. Integrability being a very stringent condition for a nonlinear PDE, one immediately realizes that the number of possible model equations which describe genuine solitons must be quite limited. A possibly even more serious limitation however, is the fact that genuine integrability (in the sense of  complete stability for {\em all} possible interactions between all soliton modes in a given model) appears to be limited to nonlinear PDEs in 1+1 or 2+1 dimensions.\footnote{The well-known connection \cite{Szimi1993, Ablo2003} of many 1+1 and 2+1 dimensional soliton equations with the self dual Yang-Mills (SDYM) equations not withstanding: although the SDYM equations contain most of the known traditional soliton systems as special reductions, no 3+1 dimensional solitonic solutions to SDYM seem to be known.}

Since integrability therefore seems far too stringent as a requirement, one is forced to conclude that genuine solitons (in the strict, mathematical, sense) are perhaps not the best candidates for the $u$ wave that describes physical particles in the double solution program. There is however still the possibility   that special solitary wave solutions to some suitably chosen nonlinear wave equation could play that role. Especially if this equation possesses certain nonlocal features, which for example would make that sufficiently narrow (peaked) solitary waves never quite collide, for example by scattering before collision. The notion of nonlocality in classical nonlinear waves equations is discussed in section \ref{RW-unstable}.

\subsection{A classical no-go theorem}\label{RW-nogo}
First it is necessary, however, to dispel a common misconception regarding the stability of solitary waves in higher dimensional models. As pointed out above, genuine solitonic interactions seem to be a phenomenon that is limited to 1+1 or 2+1 dimensional nonlinear wave equations. Traveling wave solutions, however, exist for large classes of nonlinear systems, in any dimension. For reasons that are not immediately clear to these authors, it seems that for the past 50 years a sizeable part of the physics community has interpreted a result by Derrick  \cite{Derrick}, as definite proof that no stable localized wave solutions can exist for higher dimensional nonlinear wave equations that arise from a variational principle and that possess some sort of Galileian invariance. However,  the class of variational problems for which the mathematical argument in \cite{Derrick} is correct, turns out to be much narrower than its author intended. Namely: if the wave equation one is studying also conserves the norm of the solution, then its energy is actually bounded because of norm preservation, and stable localized solutions do exist! In Appendix C we explain the problem with Derrick's argument in detail on the example of the nonlinear Sch\"rodinger (NLS) equation, a paradigm of solitonic evolution, for which the original argument of \cite{Derrick} -- when taken at face value -- mistakenly predicts that the solitons must be `unstable'.

\subsection{Rogue waves vs. solitary waves as de Broglie's $u$ waves}\label{RW-rogue}
It is important to stress that the appearance of solitons in, say the NLS equation (equation \eqref{NLS} in the appendix), or of the ground state in the Schr\"odinger-Newton equation \eqref{NS} discussed in Appendix B, is always due to the same {\em deterministic } mechanism: an initial condition radiating excess energy (and in the integrable case, possibly breaking up in the process) and thereby relaxing to a stable state. Contrary to what one is accustomed to in quantum theory, there is no `collapse' or quantum jump that needs to be postulated as some deus ex machina to obtain a reduced system. The relaxation process described above is completely deterministic and is intrinsic to the evolution described by the nonlinear wave equation one is dealing with. In the context of nonlinear extensions of quantum theory this is quite an appealing feature, since a deterministic equation with such a property can be interpreted as a model that incorporates a spontaneous collapse process.

It is in this context that the concept of a rogue wave might play an important role. Rogue waves are often defined as waves with extreme amplitudes, appearing suddenly and apparently spontaneously, as if out of nothing, just to disappear as suddenly as they arose. Adhering to this definition for now, numerous well-established testimonies  attest to the existence of oceanic rogue waves of which a deep understanding  (see \cite{Kharif}) is still lacking today. Recently however, decisive progress in the field has been realized in relation to nonlinear optics, in which optical rogue waves have been produced and observed repeatedly in a laboratory setting \cite{opticalrogue}. 
One could of course wonder whether this concept of a rogue wave, as an emergent and in most cases ephemeral phenomenon, is related to that of a solitary wave, which is inherently a persistent feature of the dynamics in a nonlinear model.\footnote{We shall, for the time being, leave aside the possibility of `rogue waves' that would emerge spontaneously but persist afterwards, for example because of soliton resonances. Such resonances will be described in section \ref{RW-unstable}.} As mentioned above, solitary waves are stable, localized, wave packets that emerge from initial perturbations (or wave fronts) that are most often of comparable size to the emerging wave. Rogue waves, on the contrary, typically arise from much smaller initial perturbations, sometimes even from mere noise.

An interesting result in this respect is reported in \cite{Hammani2010}, where numerical simulations of a nonlinear system with third order dispersion show that incoherence in the system can induce three different rogue wave regimes. If the incoherence in the system is small, then a large amplitude solitary wave emerges from small initial fluctuations and persists asymptotically, i.e.: it is stable even when interacting with the small scale fluctuations that remain in the system. If incoherence increases, there is an intermediate regime for which such a solitary wave appears and disappears intermittently, and finally when incoherence is too high, very short-lived rogue waves can appear, but only sporadically. It is reported in \cite{Hammani2010} that the statistics (in time) of these rogue wave events differs considerably from that of the intermittent behaviour in the intermediate regime. This leads to the conclusion that the amount of noise in a nonlinear system might actually determine the type of event that can take place.

Noise is of course known to drastically change the dynamics in many nonlinear systems (see e.g. \cite{Flessas2004} for a review). For the NLS equation for example, although the stability of solitons in the one (space) dimensional case is unchanged, when in two or three space dimensions additive or multiplicative noise  is added to the system, {\em any} solution will suffer a blow-up in finite time \cite{Flessas2004,Bouard1999}. Thus, in higher dimensions, in the presence of noise, sharply peaked `rogue waves' that only appear sporadically, might very well be indistinguishable from solutions that blow up because of a singularity appearing in finite time.

\begin{figure}[h]\vskip.5cm
\begin{center}
\resizebox{12cm}{!}{\includegraphics[]{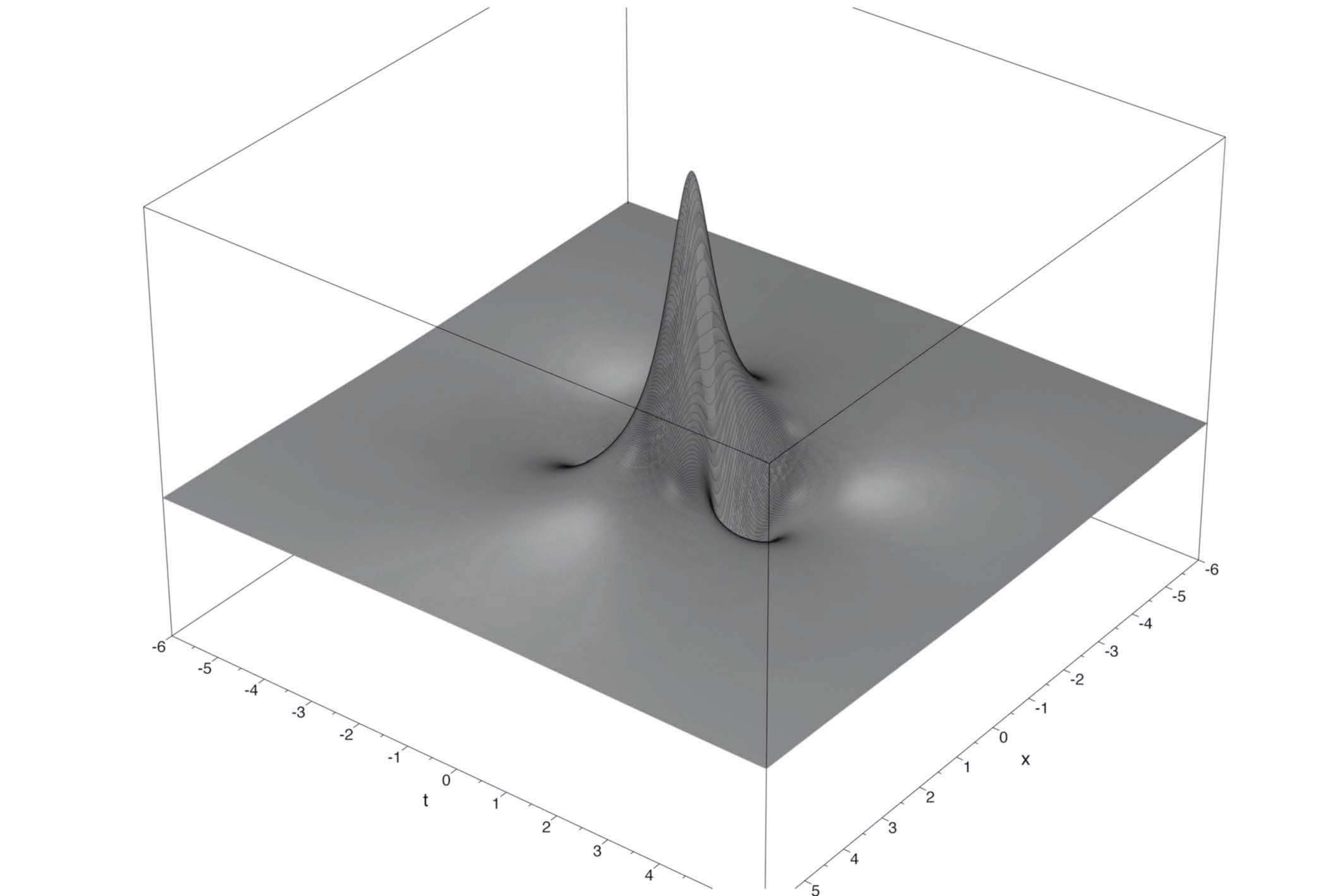}}\end{center}
\caption{Three dimensional plot of the Peregrine soliton \eqref{Pere} in the $(\tau,\xi)$ plane, clearly showing both the finite spatial and finite temporal extent of the wave.}\label{Pereplot}
\end{figure}

Even in cases where the amplitude of the rogue wave is too small for it to be considered a `singularity' that developed in a solution to some nonlinear wave equation, it seems undeniable that rogue waves are in many cases indeed related to such singularities. A popular technique for describing rogue waves for a host of nonlinear, mainly integrable, equations is as rational solutions to these equations \cite{Akhmediev}. Such rational solutions are often obtained as special limits of soliton solutions or, more generally, as degenerate cases of algebro-geometric solutions for integrable PDEs (see e.g. \cite{Matveev2016} for a very general theory of such rogue wave solutions). Perhaps the most famous example is the so-called Peregrine soliton
\begin{equation}
\big| \psi(\xi,\tau)\big| \,\propto\, \Big|\dfrac{4(1 + 2 i \xi)}{1+ 4 \tau^2 + 4 \xi^2 }-1\Big|\label{Pere}
\end{equation}
which is considered to be the basic rational solution to the self-focusing NLS equation \cite{Kibler2010}:
\begin{equation}
i\dfrac{\partial \psi}{\partial\xi} + \frac{1}{2}\frac{\partial^2\psi}{\partial\tau^2}+ |\psi|^2\psi=0.\label{sfNLS}
\end{equation}
This rational solution is of course non-singular for real $\xi$ and $\tau$, i.e. as a solution to the physical self- focusing NLS equation, but it clearly does have a singularity locus in the complex $(\tau,\xi)$ plane: the rational curve $\xi^2+\tau^2=-1/4$. Hence, the sudden appearance and subsequent disappearance of the Peregrine soliton (depicted in figure \ref{Pereplot}) can be understood as the result of the closest approach of this solution in the {\em real} $(\xi,\tau)$ plane to this singularity locus. The closest approach obviously occurs at $(\xi,\tau)=(0,0)$, which corresponds to the maximal amplitude of the Peregrine soliton in figure \ref{Pereplot}.

Traveling wave-reductions of 1+1 dimensional nonlinear PDEs, generically, have movable singularities in the complex plane (i.e., singularities the positions of which are not fixed by the coefficients of the equation, but which depend on the initial conditions). Hence, for a given nonlinear PDE, some of its traveling wave solutions will -- in {\em real} time and space -- necessarily pass `near' a complex movable singularity, the position of which depends on the initial conditions one chooses. For all practical purposes, such solutions will therefore exhibit rogue wave behaviour. Whether this type of behaviour is related to the blow-up phenomenon for certain nonlinear PDEs is not entirely clear, but it does seem to be at the heart of the rogue wave phenomenon. This being said, since general solutions to such traveling wave reductions of nonlinear PDEs will, generically, involve Painlev\'e transcendents, they cannot be expressed in closed form. Hence, only very special solutions will be representable in analytic form, and algebro-geometric solutions (involving theta functions) and their degenerations such as rational solutions therefore constitute an important class of examples.

Although such rational solutions are very special, they do appear to have some bearing on the rogue wave phenomenon as observed in the real world. It took a considerable amount of time before the existence of the Peregrine soliton in a physical (optical) system (described by the self-focusing NLS) could be experimentally confirmed \cite{Kibler2010}, but since then it has been observed, for example, in multicomponent negative ion plasmas \cite{Bailung2011} and in experimental water-wave tanks \cite{Chabchoub2012wt}. Moreover, higher-order Peregrine solitons (e.g. degenerations to rational solutions of multi-soliton solutions) have also been observed \cite{Chabchoub2012ho}. Indeed, a rather popular technique for simulating rogue waves in experimental situations is to start from well-defined initial conditions, more or less tailored to match very specific exact solutions to the model equations under consideration (e.g. rational solutions in the NLS case  or special shaped line solitons in the case of the Kadomtsev-Petviashvili (KP) equation \cite{Kodama}). Recent numerical results \cite{Kao} show that, at least in the KP case, i.e. for rogue waves in shallow water, the solitonic approach fits rather well with the initial value problem for selected initial profiles. However, contrary to what is known for the one dimensional NLS equation \cite{Kibler2011}, it is not clear yet whether these candidates really fit with the experimentally observed rogue waves. In other words, it is unclear whether they can be connected to generic initial conditions for the nonlinear PDEs of interest, especially in the presence of noise. 

A more fundamental problem however -- if one thinks of applications to quantum theory -- is that there does not seem to be a valid analytic description yet of rogue waves in 3+1 dimensions. This problem is of course mainly due to the aforementioned dearth of integrable systems in higher dimensions. Furthermore, so far, no family of special solutions to NLS (or any other nonlinear wave equation for that matter) has been identified that would exhibit the sort of transition between different regimes that has been observed in \cite{Hammani2010} and it is not clear whether the behaviours in these three regimes are due to the same phenomenon or rather, whether solitary wave-type (i.e. persistent) behaviour is actually fundamentally different from (sporadic) rogue-wave like behaviour. It is therefore still possible that, one day, the difference between `rogue' and `solitary' waves turns out to be a mere semantic one, i.e. that they are essentially the same phenomenon but simply with relaxation time scales (and frequencies of occurrence) that differ according to the physical circumstances. For the time being it seems best however to treat them as separate phenomena, where the term `rogue wave' refers to a short-lived (ephemeral) localized wave appearing and disappearing sporadically, and where a `traveling' or `solitary' wave is a localized wave that persists after it was created (or introduced as an initial condition) in the nonlinear evolution. In this sense, rogue waves might be useful in the context of deterministic models for spontaneous collapse (cf. appendix A and B) but in the context of de Broglie's double wave program it is the solitary wave concept that seems more relevant.

\subsection{Nonlocality and instability in nonlinear wave equations}\label{RW-unstable}
As mentioned before, if de Broglie's $u$ wave is to be interpreted as a solitary wave, then this wave will almost certainly be a solution to a non-integrable nonlinear wave equation. In which case this equation must contain sufficiently many `nonlocal' features, if one wants to avoid problems with colliding solitary waves in a description of a many particle system. No definite candidate for such a model equation has been found yet, but in \cite{durt-single}, section 2 (see also \cite{durt-epl} and \cite{CDW}), some plausible {candidates are investigated}.

`Nonlocality' in a nonlinear PDE comes in many forms. The Schr\"odinger-Newton equation (cf. equation \eqref{NS} in Appendix B) is manifestly nonlocal because of its {integro-differential} term, but its purely differential formulation
\begin{eqnarray}
i \hbar\dfrac{\partial\Psi(t,{\bf x})}{\partial t}=-\dfrac{\hbar^2}{2m}\Delta\Psi(t,{\bf x})
+ m V(t,{\bf x}) ~\Psi(t,{\bf x})\nonumber\\
\Delta V = 4 \pi G m \big|\psi(t,{\bf x})\big|^2,\label{NSdiff}
\end{eqnarray}
is of course equally nonlocal because of the coupling of the $\psi$ and $V$ fields through a Poisson equation. Other interesting examples of systems with this type of nonlocality are the generalized Manakov system 
\begin{eqnarray}
i a_t + a_{xx} \pm \lambda |a|^2 a + 2 a (\eta+\zeta)=0\,,\nonumber\\ \eta_{tt}-c_L^2 \eta_{xx} = - \mu_L(|a|^2)_{xx}\,,~~ \zeta_{tt}-c_T^2 \zeta_{xx} = - \mu_T(|a|^2)_{xx}\label{gZ},
\end{eqnarray}
which arises in the description of nonlinear acoustic surface waves in crystals \cite{Maugin1999}, or the system
\begin{equation}
\left( \partial_x^2 + \partial_y^2 \right) A + 2 i k\, \partial_z A + \frac{2k^2}{\eta_0}\,\eta\,A=0\,,\qquad \Delta \eta = - \tilde\kappa\,\big| A\big|^2,\label{Rot}
\end{equation}
which is a paraxial nonlinear wave equation that describes light in a medium with an optical thermal nonlinearity that induces a change in the refractive index $\eta$ of the medium \cite{Rot2006}. 

As described in \cite{Maugin1999} (pp. 213), although non-integrable\footnote{The reduction $\lambda=\mu_T=0, \xi(x,t)\equiv0$ of \eqref{gZ} is the usual Zakharov system which is a known integrable system, integrable through inverse scattering techniques.}, system \eqref{gZ} has solitary wave type solutions that possess a remarkable and in the present context highly desirable property: above a certain threshold velocity, collisions of solitary waves are impossible because the solitary waves effectively `brake' before colliding and practically come to rest at a finite distance from each other. This sort of behaviour is of course only possible because of the nonlocal character of the interaction.

The dynamics described by system \eqref{Rot} have even more spectacular nonlocal features. As described in \cite{Rot2006}, the (optical) solitary waves that arise in the medium described by \eqref{Rot} can be shown, experimentally, to attract each other even when their optical fields do not overlap. A three dimensional experimental set-up even led to observations of solitary waves capturing each other in a spiralling motion on a circular orbit, with tangential velocities that are in fact independent of the distance between the solitary waves. 

However, nonlinear wave equations do not always need to have an explicit nonlocal term in order to exhibit nonlocal features in their solutions. In fact, because of the global nature of  the conservation laws for such wave equations, it is possible for an interesting interplay between nonlinearity and nonlocality to manifest itself. As is well known, classical field equations that can be derived from a variational principle with a symmetry group with a finite (or countably infinite) number of generators\footnote{It might be less well-known that variational problems with symmetries with a non-countable number of infinitesimal generators, like those arising in general relativity or general gauge theories, might not have proper conservation laws at all. This is essentially the content of Noether's second theorem \cite{Sid}.} have local conservation laws of the form $\rho_t = {\rm div} F$, and hence their conserved quantities have to be expressed as integrals over phase space : $\DIS \frac{\partial}{\partial t}\int_\Omega\, \rho\,dv = 0$.

In the case of linear or linearizable systems, for well separated states, the superposition principle still guarantees some sort of `approximate' local conservation of such conserved quantities. For example, although the energy or momentum of such a system is only ever given exactly by an integral over the full phase space $\Omega$, as long as two localized waves in a linear system are well separated, their respective energies and momenta will approximately be conserved locally, over the extent of each wave separately. This approximate view would allow us to consider each solitary wave as an individual entity in such a system.
On the other hand, for nonlinear systems, such an approximate local conservation is not guaranteed at all.
For example, take a situation where two solitary waves evolve according to a nonlinear evolution rule, at a considerable distance from each other. In the absence of a superposition principle such a two-hump state will not be an exact solution to the nonlinear equation and it will, over time, relax to a stable state through some radiative process. Now, since the conservation laws for such a system only dictate global invariance, over the whole of $\Omega$, locally almost anything can happen: one of the solitary waves might for example radiate all or most of `its locally visible' mass and energy out into the rest of phase space and suddenly disappear, or at least change its amplitude considerably. Moreover, if there is no maximum speed (a `speed of light') in the system, the radiative part of the solution that is generated in this relaxation process will permeate the entire phase space instantaneously. Which means that in such a system, at least in principle, there will be some spooky action at a distance, as this radiation perturbs the other solitary wave instantaneously. 

For example, in \cite{spooky}, an extremely weak and long-range interaction between optical cavity solitons in an optical fibre is described. Since these are interactions in a real physical system, their speed is of course limited by the speed of light, but even then it was reported that the acoustic waves in the fibre responsible for the interaction were able to influence solitons separated by an effective distance of the order of an astronomical unit.

\begin{figure}[H]\vskip0cm
\begin{center}
\resizebox{12cm}{!}{\includegraphics{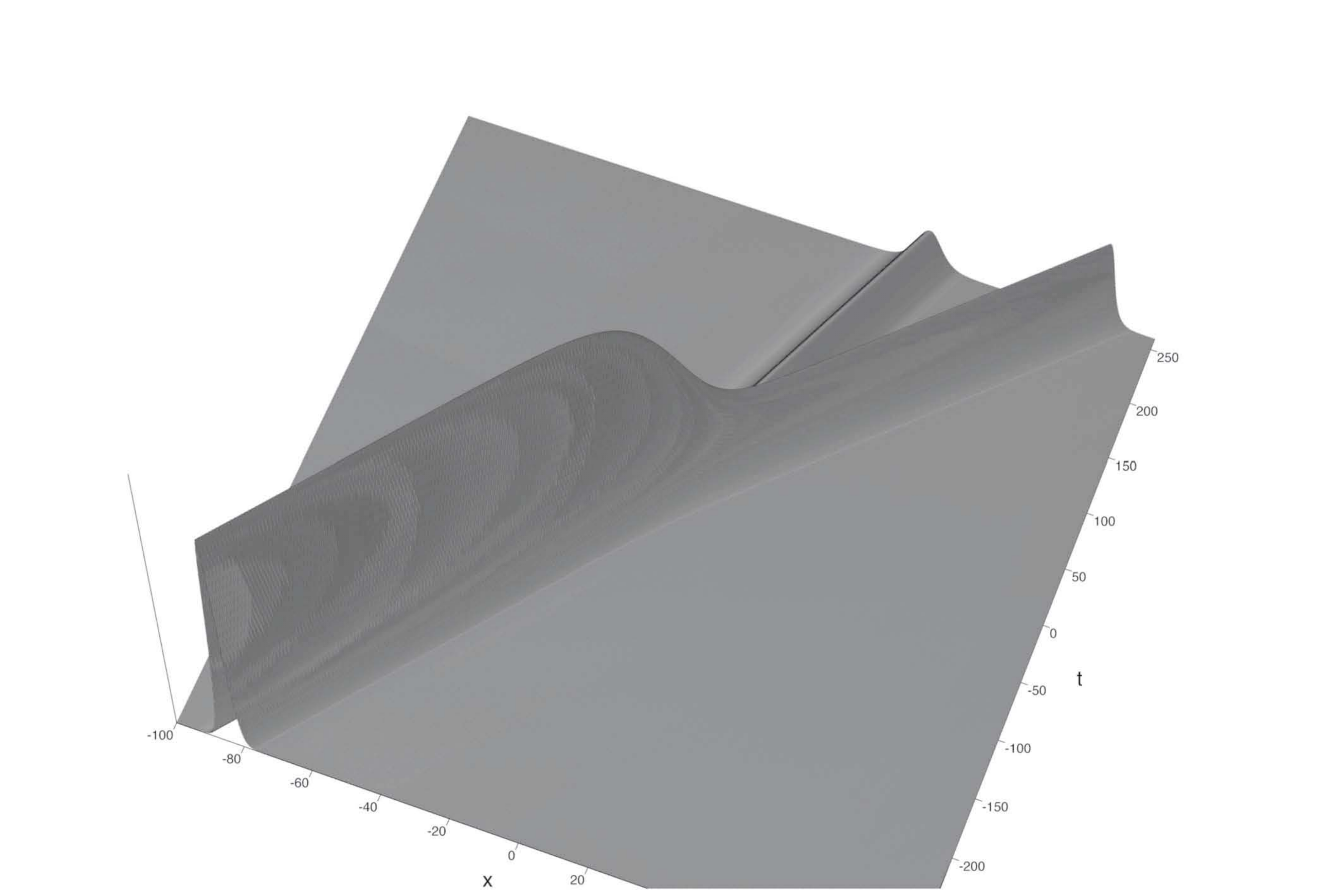}}
\end{center}
\caption{A plot in the $xt-$plane of the resonant solitary wave $4\frac{\partial^2}{\partial x^2}\log\big(1+e^{k^3t-kx}+e^{[1+3k(k-1)] t -x +\delta}\,\big)$ with amplitude 1 and velocity $1+3k(k-1)$, which decays into two smaller but stable solitary waves (with amplitudes $k$ and $(1-k)$, and speeds $k^2$ and $(1-k)^2$ respectively) for $k=0.6$ and internal phase constant $\delta=0$. Time runs from t=-300 to 260.}\label{resonance}
\end{figure}

Another feature of nonlinear wave equations which plays an important role in manifestations of such `implicit' nonlocality\footnote{As opposed to the `explicit' nonlocality of equations (\ref{NSdiff}--\ref{Rot}).}, is that of so-called resonant solutions \cite{Taj1982,Lambert1988}. These are solitary wave solutions to certain nonlinear (not necessarily integrable) wave equations that are intrinsically unstable in the sense that they either `decay' into several stable solitary waves or, conversely, merge into a solitary wave that is asymptotically stable. The exact instant in time however where such fission or fusion of solitary waves will take place is governed by an internal phase of the solution, and is therefore in principle unknowable to an outside observer. The existence of such solitary waves in atomic nonlinear chains has been demonstrated in \cite{Flytz1987}.
An example of a resonant solution to the equation discussed in \cite{Lambert1988} is shown in figure \ref{resonance}. 

Decaying resonant solutions are extremely interesting in the context of nonlocality because of their intrinsic instability. Since the internal phase that determines the instant at which the resonant state will decay is, in general, strongly influenced by the interactions with other waves in the system, even the tiniest perturbation of this metastable state, in the most remote part of phase space, might provoke an instant decay, as it were resetting the internal clock of the resonant wave. Moreover, as will often be the case, if the system is non-relativistic then this influence will be instantaneous, offering a dramatic example of a counter intuitive nonlocal effect (in a classical, non-quantum, sense), solely due to the nonlinearity of the system.

\section{No-go theorems: nonlinearity vs. nonlocality and no-signaling\label{nogo1}}
\subsection{Gisin's no-go theorem.\label{nogo2}}Several authors (Polchinski \cite{polchinski}, Gisin \cite{gisin}, Czachor \cite{czachor} and others) have shown that nonlinear generalisations of the Schr\"odinger equation, at the level of the configuration space, are generically nonlocal. Roughly summarized, Gisin's argument \cite{gisin} goes as follows: nonlinear corrections to the linear Schr\"odinger equation make it possible, in principle, to distinguish different realizations of the same density matrix (this is related to the so-called mobility property  \cite{czachor}, made explicit by Mielnik \cite{mielnikmobility}). By performing a local measurement on a system A that is entangled with a distant system B, one is able, by collapsing the full wave function, to obtain realizations of the reduced density matrix of the system B which differ according to the choice of the measurement basis made in the region A. Therefore, in principle, nonlinearity can be  a tool for sending classical information faster than light, contradicting the no-signaling property valid in the framework of linear quantum mechanics \cite{gisin}.

These nonlocal features have been used (Christensen and Mattuck \cite{christmatt}, Durt \cite{durt})  to explain the violation of Bell inequalities in the framework of the so-called Bohm-Bub model \cite{bohmbub}, which is a stochastic hidden variable model aimed at simulating the collapse process thanks to a nonlinear modification of the linear quantum theory. Actually, the Bohm-bub model is a prototype of what are nowadays called collapse models, which aim at providing a realistic description of quantum phenomena and in particular of the collapse process, which in this context,  is assumed to result from the presence of an intrinsic nonlinear modification of the Schr\"odinger evolution.

As has been shown by Gisin, a large class of collapse models (e.g. the collapse models discussed in  \cite{GRW,pearle,Bassi,diosi,Adler,flash}) as well as a more standard interpretation of quantum theory in which a von Neumann collapse occurs during the measurement process, in accordance with the Born rule, in fact do not contradict the no-signaling property thanks to the combination of stochasticity and nonlinearity in these models, which somehow conspires so as to respect causality ``on average'' \cite{gisin2}.  This property is explicitly addressed in appendix D  where the relation between the Born rule and no-signaling is discussed.

\subsection{Realism and nonlocality\label{nogoreal}}
The question of nonlocality, as well as the question of relativistic invariance are problematic in {\em all} realistic interpretations  of the quantum theory. For instance, in an apparently Lorentz covariant collapse model (cf. appendix A) -- the so-called flash ontology proposed by Tumulka \cite{flash} -- it is assumed that spontaneous localizations are distributed according to a Lorentz-covariant distribution, when computed in different inertial frames. However, as noted in \cite{flashcomment}, one is always required in this model to select a privileged frame to begin with. Once such a frame is selected, Tumulka's model makes it possible to compute the influence of quantum jumps (flashes) on the wave function assigned to the system (here $N$ non-interacting particles obeying Dirac's equation), but this aspect of the model is manifestly not Lorentz covariant, although the resulting distribution of quantum jumps (flashes or events in space-time) {\em is} Lorentz covariant. This is a general feature of dB-B and collapse models (and more generally of all collapse models, as explained in  \cite{mielnikbottle}): a privileged frame must be chosen in which nonlocality is necessarily present. 

As mentioned above, the no-signaling condition is respected in all these models thanks to a quasi-miraculous conspiracy of stochasticity and nonlocality, which results in the disappearance of nonlocal action-at-a-distance after it has been averaged in accordance with the Born rule.\footnote{This is also true for what concerns less realistic theories in which the collapse process is induced by the measuring apparatus: the no-signaling theorem is valid provided the Born rule is obeyed. We shall come back to this point in the Appendix D.}

It is however generally admitted that Gisin's no-go theorem does hold in the case of purely deterministic generalisations of the linear Schr\"odinger equation, such as the aforementioned Schr\"odinger-Newton equation \eqref{NS}  \cite{guignol,guignol2}  or the NLS equation, for the simple reason that in the case of deterministic evolution equations, stochasticity is in principle absent. This being said, as we shall discuss next,  there do exist several different ways to interpret Gisin's result.

 \subsection{Circumventing Gisin's theorem (1): nonlinearity without nonlocality}
 Gisin's argument relies on a nonlinear modification of the Schr\"odinger equation due to Weinberg \cite{weinberg1} {(cf. section \ref{secwein})}. Now, it could be that there exist specific nonlinear modifications of the Schr\"odinger equation that do not belong to the class considered by Gisin, but no such modification is known. For instance, Bilianicki-Birula has proposed a nonlinear potential {\cite{bbmy,bb}} that factorizes into local potentials whenever the full wave function factorizes into a product of local wave functions. This property is not sufficient however to escape the limitations imposed by Gisin's theorem if the subsytems are entangled.  Another example is provided by the so-called Hartree regime, which is valid precisely in situations where no entanglement appears {between} the subsystems $A$ and $B$, as  discussed in \cite{zeit}. In theorem 3 of that reference,  the following is shown:

A bipartite system, initially prepared in a factorisable (non-necessarily pure) state 
$$\rho_{AB}(t = 0) = \left(\rho_A(t = 0) \otimes \rho_B(t = 0)\right),$$
remains in a factorisable state throughout the evolution, 
$$\rho_{AB}(t = 0) = (\rho_A(t = 0) \otimes \rho_B(t = 0))
\quad\forall t\geq 0,$$
if and only if the effect of the Hamiltonian can be factorised as follows:
\begin{eqnarray*}
\forall t\geq 0,\quad [H_{AB} (t),\rho_{AB} (t)]\\=([H^{eff}_A(t),\rho_A(t)])\otimes \rho_B(t)+\rho_A(t)\otimes([H^{eff}_B.(t),\rho (t)]),
\end{eqnarray*}
 where $[H^{eff}_A(t),\rho_A(t)]=Tr_B([H_{AB} (t),\rho_{AB} (t)])$ and $[H^{eff}_B(t),\rho_B(t)]=Tr_A([H_{AB} (t),\rho_{AB} (t)])$, {in which case $i\hbar {\partial \over \partial t} \rho_A(t)=[H^{eff}_A(t),\rho_A(t)]$ and $i\hbar {\partial \over \partial t} \rho_B(t)=[H^{eff}_B(t),\rho_B(t)]$. This constitutes what is called the mean field or Hartree approximation in which any subsystem ``feels'' the average potential generated by the other subsystems. Obviously,  the reduced dynamics of the subsystems $A$ and $B$ is highly nonlinear because the local, effective, dynamics of one subsystem contains averages performed on the second subsystem.} Contrary to Gisin's argument however, the ensuing dynamics does not make it possible to distinguish different realizations of  a same initial density matrix $\rho_{AB}(t = 0) = \rho_A(t = 0) \otimes \rho_B(t = 0)$, as the dynamics is of first order in time, and formulated only in terms of the density matrix $\rho_{AB}(t ) = \rho_A(t) \otimes \rho_B(t )$.
 
  \subsection{Circumventing Gisin's theorem (2): Chaotic dynamics, equilibrium distribution and onset of the Born rule\label{nogoBorn}}One of the most interesting aspects of de Broglie's pilot-wave theory is the possible existence of so-called quantum non-equilibrium distributions \cite{valentini-phd}, which are distributions in which the configurations (positions or fields) are not distributed according to Born's law (e.g., $\rho(t,{\bf x})\neq|\psi(t,{\bf x})|^2$ for a single particle system).  Results from experiments realized on such systems would obviously disagree with standard quantum mechanics. Although there is no doubt that such systems are in principle allowed in the pilot-wave theory, one has to admit that none have been observed to this day (as all experiments realized on quantum systems yield results that are in perfect agreement with Born's law). Therefore, if quantum non-equilibrium distributions existed in the early stages of the universe, they must have quickly relaxed to quantum equilibrium. This process of relaxation to quantum equilibrium is supported by numerous numerical simulations  \cite{valentini042,cost10,toruva,colin2012} and the process of relaxation has been thoroughly investigated within chaos theory in \cite{efthymiopoulos}. Despite the fact that dB-B theory is a no-collapse theory -- thus prohibiting the use of arguments such as those used in the Appendix where the collapse process is invoked -- it is well-known that it does respect the no-signaling requirement, provided equilibrium has been achieved. Once again, no-signaling results in this case from of a perfect balance between stochasticity ({\it \`a la} Born) and signaling.

All these results suggest a possible route for circumventing Gisin's theorem: if, in accordance with the de Broglie double solution program, particles are identified with solitonic solutions of a nonlinear modification of Schr\"odinger's equation such that the trajectories of these solitons obey dB-B dynamics, then, making use of the aforementioned results concerning the Born rule and no-signaling (section \ref{nogo2} and appendix D), the no-signaling condition is respected once the process of relaxation is achieved.     
  
 \subsection{Not circumventing Gisin's theorem (3): supraluminal telegraphy\label{nogomust}} Another interesting line of research is that where one accepts the implications of Gisin's theorem, for instance in the framework of semi-classical gravity models {\it \`a la} Diosi-Penrose, which leads to the question of when and where nonlocal, supraluminal communication is possible.     It is worth mentioning that several experiments were performed by Gisin's group in the past in order to reveal hypothetical finite speed influences (spooky actions at a distance) that might occur during the collapse process. These experiments were realized with long-distance entangled pairs of photons \cite{gisinspooky} and they all led to negative results, confirming thereby the standard predictions of the quantum theory (see also \cite{longdistancesatellite}  for proposals of similar tests to be realized in satellites\footnote{It is also well-known in the context of general relativity that, in principle, Lorentz invariance is broken by the gravitational interaction \cite{longdistancesatellite}. However at the surface of the Earth, no violation of Lorentz symmetry has been measured so far. In CERN for instance the standard model has been repeatedly confirmed, which confirms indirectly that local physics can be explained by relativistic quantum field theory. In Gisin's lab in Geneva, where violations of Bell's inequalities have been realized with pairs of entangled photons separated by a spacelike distance of 10 km, the quest for a privileged reference frame has also systematically led to negative results \cite{gisinspooky}. }). In the past \cite{durt}, it has been shown however, that certain nonlinear modifications of Schr\"odinger's dynamics (e.g. the Bohm-Bub dynamics) make it possible to reproduce a large set of experimental data, even when entangled systems are involved. To our knowlegde, no systematic study of nonlinearity induced signaling has been performed up to now. {For instance, the aforementioned negative result experiment \cite{gisinspooky} certainly imposes experimental bounds on nonlinearity induced signaling, but it is not clear yet to which extent the various nonlinear extensions of the Schr\"odinger equation mentioned in that paper have been ruled out by this type of experiment.}

\section{Discussion and conclusion}
\subsection{de Broglie versus Bohm} This review paper, sketching the context and relevance of de Broglie's double solution program in present-day physics, led us to discuss several properties that are usually not considered in the framework of the so-called causal interpretation of de Broglie-Bohm. 

This is because most often de Broglie's original ideas were forgotten, as they disappeared in the simplified version of the double solution program that is the pilot wave theory 
(see \cite{fargue-proc}). In the pilot wave formulation, the mechanism through which the real wave and the $\Psi$ wave interact is silenced, and in the Bohmian approach, the $u$ wave is replaced by a material point. 

{This simplifying approach however invites certain important criticisms. For example, the mere idea of a material point violates what is termed the `No Singularity Principle' in \cite{gouesbet}, according to which 

{\it ``[L]ocal infinity in physics is not admissible, that is to say \ldots nature (locally) abhors infinity \ldots"}. 

Material points (as well as the singular solutions of de Broglie's double solution program) of course violate this principle, but this is obviously not the case for hump-type, non-singular solutions of de Broglie's program which are characterized by a finite-but-non-zero size and a finite amplitude. } 

Whereas in the Bohmian approach the linear structure of the dynamical equation is respected, our analysis shows that important features of the dB-B dynamics such as nonlocality and nonseparability can be shown to derive from the assumption that the evolution of the system is governed by fundamentally nonlinear PDEs.

Last but not least, as noted by Bush \cite{bush}, the physics of `walkers' (outlined in the next paragraph) is reminiscent of de Broglie's double solution program, rather than of the dB-B interpretation, since in hydrodynamics, at the macroscopic level, only continuous and non-singular quantities can define the state of the system: the physics of walkers is wave-monistic.
\subsection{Walkers: a macroscopic realization of the double solution program?}Even in the field of quantum wave mechanics, de Broglie-Bohm (dB-B) trajectories \cite{debroglie60,bohm521,Holland} remained for years a rather confidential and academic topic\footnote{Although dB-B trajectories \cite{debroglie60,bohm521,Holland} have been studied in relation with the measurement problem, they are often considered as conceptual tools rather than empirical realities, among other reasons because the dB-B dynamics is often considered to be an ad hoc reinterpretation of the standard quantum theory.}. Recently however, they have regained a certain prominence since they were realized in a laboratory setting in artificial macroscopic systems, as bouncing oil droplets or so-called walkers \cite{couder1,couder2,couder3,bush}. These take the form of oil droplets bouncing off  the surface of a vibrating bath of oil, excited in a Faraday resonance regime (the walkers are prevented from coalescing into the bath, the vibration of which creates a thin film of air between its surface and the droplet, and therefore seem to levitate above it). 

Walkers exhibit rich and intriguing properties. Roughly summarized, they were shown experimentally to follow dB-B-like quantum trajectories.  For instance, when the walker passes through one slit of a two-slit device, it undergoes the influence of its ``pilot-wave'' passing through the other slit, in such a way that, after averaging over many dB-B like trajectories, the interference pattern typical of a double-slit experiment is restored and this despite the fact that each walker passes through only one slit. The average trajectories of the drops exhibit several other quantum features such as orbit quantization, quantum tunneling, single-slit diffraction, the Zeeman effect and so on. Another surprising feature is a  pseudo-gravitational interaction that has  been observed between two droplets. In \cite{couder2}, for instance, it is mentioned that: 

{\it [D]epending on the value of d [which represents the impact parameter of the collision] the interaction is either repulsive or attractive. When repulsive, the drops follow two approximately hyperbolic trajectories. When attractive, there is usually a mutual capture of the two walkers into an orbital motion similar to that of twin stars  \ldots} .

These observations suggest that a `fluidic', hydrodynamical formulation of wave mechanics is possible, in which the droplet would play the role of de Broglie's $u$ wave, while the properties of the environmental bath are assigned to the $\Psi$ wave of de Broglie.
\subsection{Conclusion.}
Nearly one century after their genesis, de Broglie's ideas remain largely unexplored. To this day, only a few serious attempts have been realized in order to reformulate quantum dynamics as a nonlinear dynamics. This approach might, however, constitute an interesting bridge towards quantum formulations of gravitation\footnote{{Penrose argued that \cite{penrose2} 
{\it [T]he case for ``gravitizing'' quantum theory is at least as strong as that for quantizing gravity \ldots }, 
an idea to be compared to M{\o}ller and Rosenfeld's views on quantum gravity (cf. appendix B).}} {which has the merit that it draws from general relativity the lesson that nonlinearity and gravitation are indissociable. It also has the merit that it might explain, in gravitational terms, the cohesion of elementary particles (and, to a larger extent, of matter as a whole), a problem Einstein already faced when he developed the theory of the photon (cf. the quote in the introduction) and which was originally addressed by Poincar\'e himself in 1905 when he introduced the so-called Poincar\'e pressure in order to explain why electrons, viewed as force fields, do not spread with time.} This approach also forces us to adopt a rich and colorful description of Nature in which typical quantum features such as nonlocality and interdependence share the stage with typical features of complex classical systems such as unpredictability and nonlinearity. 
This might be the price to pay if one wishes, following de Broglie, to develop a non-dualistic representation of the physical world. 

We hope that these ideas will ultimately lead to new predictions that could be tested in the laboratory, and bring these `old' debates about the measurement problem and the interpretation of the quantum theory back into the realm of experimental physics.   

\section*{Acknowledgements}
This work was made possible thanks to the financial support of a grant from the FQXi Foundation (Quantum Rogue Waves as Emerging Quantum Events-FQXI project FQXi-RFP-1506).
S. Colin thanks J. Theiss (Theiss Research) for his dedicated work in administering part of that grant.
This work also benefited from the support of grants 21326 and 60230 from the John Templeton Foundation. T. Durt thanks J. Uffink for bringing to his attention the paper of M. Born \cite{born} 
(private communication, Banff, August 2016) .

\appendix
\section{Collapse models}\label{collapse_app}
The main idea behind collapse models is to replace the deterministic and linear Schr\"odinger equation 
by a stochastic and nonlinear one. This approach was pioneered by Pearle \cite{pearle0} and Ghirardi, Rimini and Weber \cite{GRW} (whose model is commonly referred to as the GRW model). 
The aim of these models is to solve the measurement problem by explaining why macroscopic objects are localized.

The basic ingredient of the GRW model is a hypothetical mechanism of spontaneous localization (SL)
which would be active everywhere in our universe and which would ultimately explain why classical objects are characterized by an unambiguous localisation in space \cite{Bassi}. The GRW model \cite{GRW} predicts for instance that the quantum superposition principle is violated in such a way that a macroscopic superposition (a Schr\"odinger cat state) will collapse into a well-resolved localised wave packet (with an extent of the order of 10$^{-7}$ m) after a time inversely proportional to the mass of the pointer. This time becomes very small in the classical limit, seen here as the large mass limit. For instance, the original GRW model predicts that the collapse time is of the order of 10$^{-7}$  s, for a pointer of mass equal to 10$^{23}$ a.m.u..

Towards the end of the 1980's several refinements of this model, such as the continuous 
spontaneous localization (CSL) model (\cite{pearle} and \cite{CSL}), appeared and there now exists an extended zoology of SL models, such as those attributing the source of spontaneous localization to 
\begin{itemize}
\item a dedicated universal localization mechanism (GRW \cite{GRW}, CSL \cite{pearle,CSL}, Quantum Mechanics with Universal Position Localization (QMUPL) \cite{Bassi,diosi} and Adler's SL models \cite{Adler}), 
\item self-gravitation (Di\'osi \cite{diosi} and Penrose \cite{penrose}), 
\item fluctuations of the spacetime metric (K\'arolyh\'azy, Frenkel {\it et al.} \cite{frenkel}), 
\item quantum gravity (Ellis {\it et al.} \cite{QG1,QG2}) 
\end{itemize}
and so on. In spite of their differences, these models have in common that they lead to accurate predictions regarding the quantum-classical transition \cite{arndt-horn}. We invite the interested reader to consult the recent and exhaustive review paper of Bassi {\it et al.} dedicated to this topic \cite{Bassi} as well as reference \cite{Romero} where a careful estimate of the SL parameters assigned to these various models is provided.

\section{The Schr\"odinger-Newton equation}\label{sne-app}
In the 1960's,  M{\o}ller \cite{Moeller} and Rosenfeld \cite{Rosenfeld} proposed that the source term in Einstein's equations would be the mean quantum stress-energy tensor. The basic idea behind this proposal is that whereas matter is quantized, space-time is not. In the non-relativistic limit, this leads \cite{Diosi84,giulini2012} to the Schr\"odinger-Newton  integro-differential equation:

\begin{equation}
{i}\hbar\frac{\partial\Psi(t,{\bf x})}{\partial t}=-\hbar^2\frac{\Delta\Psi(t,{\bf x})}{2m}
-Gm^2\int {d}^3 x'\frac{\rho(t,{\bf x'})}{|{\bf x -x'}|})\Psi(t,{\bf x}),\label{NS}
\end{equation}where $\rho(t,{\bf x'})$ is the quantum density.
An immediate consequence of this equation is that even a ``free'' particle will feel its own gravitational potential, due to the source $\rho(t,{\bf x'})=|\Psi(t,{\bf x'})|^2$. In other words, the full energy now contains a contribution from
the gravitational self-energy, proportional to 
\begin{equation}
-{Gm^2\over 2}\int {d}^3 x {d}^3 x'\frac{\rho(t,{\bf x})\rho(t,{\bf x'})}{|{\bf x -x'}|}=-{Gm^2\over 2}\int {d}^3 x {d}^3 x'\frac{|\Psi(t,{\bf x})|^2 |\Psi(t,{\bf x'})|^2}{|{\bf x -x'}|},
\end{equation}
which is the average value, with weight $\rho=|\Psi|^2$, of
\begin{equation}
-{Gm^2\over 2}\int {d}^3 x'\frac{\rho(t,{\bf x'})}{|{\bf x -x'}|}~.
\end{equation}
Perhaps the most interesting feature of the Schr\"odinger-Newton equation is its essential nonlinear character. Non-linearity is of course already present at the classical level due to the fact that the gravitational field also contributes to the stress-energy tensor. At the quantum level however, nonlinearity is an intrinsic feature of the mean field approximation. This has been noted by a series of physicists (Jones \cite{Jones}, Penrose \cite{penrose} and many others), who insisted on the fact that, if we want to provide a quantum formulation of the gravitational interaction, it is inconsistent to neglect the intrinsic nonlinear nature of gravitational self-interaction.  It was also recognised very early on that gravitational self-interaction might have something to do with spontaneous localisation \cite{Diosi84}.

The Schr\"odinger-Newton equation is part of a larger class of interesting physical models called Wigner-Poisson (or quantum Vlasov-Poisson) systems that describe particle density functions in phase-space \cite{Wigner, Illner}. In this context, global existence and uniqueness of solutions to the Schr\"odinger-Newton equation has been shown by Illner and Zweifel in \cite{Illner}, where the asymptotic behaviour of its solutions for the case of a repulsive (Coulomb) potential was also discussed. As explained by these authors, the difference between the repulsive and the attractive (gravitational) cases is in fact far greater than a mere change in sign in the interaction term in the equation. Essential differences arise when one tries to discuss the stability and asymptotics for generic solutions for both types of systems. The general asymptotics and stability of solutions to the 
Schr\"odinger Newton equation \eqref{NS} is discussed by Arriola and Soler in \cite{Arriola}, refuting the widely held belief that {\it ``[A]n attractive force might lead to a blow-up in finite time \ldots"} (as claimed in \cite{Brezzi}, where the existence of solutions for the repulsive case was first proven). The stability of stationary solutions to the Schr\"odinger-Newton equation \eqref{NS} was proven much earlier by Cazenave and Lions \cite{Cazenave} and, independently, by Turitsyn \cite{Turitsyn}.
Moreover, it has been shown in \cite{Arriola} that there exist spherical symmetric solutions for the static Schr\"odinger-Newton equation corresponding to excited states with breather-like properties. Numerical evidence for such states also seems to exist (cf.\cite{Ehrhardt}). 
Furthermore, numerical results \cite{Harrison} for axially symmetric solutions to the static Schr\"odinger-Newton equation, show that these correspond to energies that are higher than that of its spherically symmetric ground state.

Another interesting result is due to Arriola and Soler \cite{Arriola}, who have shown that for negative values of the energy, the natural dispersion of a state is inhibited by the nonlinearity in \eqref{NS} and that this leads to a collapse\footnote{More precisely: in \cite{Arriola} it is shown that initial states with positive energy will expand asymptotically, and inhibition of such dispersion with an ensuing collapse is only possible for initial stationary states with negative total energy.\label{blobl}} to solutions that oscillate around the ground state of the equation. In a physical setting, this criterion yields a lower bound on the mass of self-gravitating objects of a given size below which such a collapse cannot occur. Several attempts have been made at simulating the Schr\"odinger-Newton equation \cite{Carlip, Guzman2003, Guzman2004, Giulini, Ehrhardt, vanMeter} mostly in the spherically symmetric case (although general schemes do exist, cf.\cite{Ehrhardt}) with the aim of demonstrating the existence of such a gravitational collapse. Numerical evidence presented in \cite{vanMeter} indicates that for sufficiently `massive' initial conditions (such that they have negative energy),  there is indeed a clear contraction of the initial condition to a unique and stable state.  Moreover, the author of \cite{vanMeter} gives a condition (based on some numerical estimates) for an initial gaussian wave packet to undergo such a collapse : $\sqrt{r^2}$, its spatial extent, should satisfy the inequality 
\begin{equation}\sqrt{r^2}\leq (1.14)^3(\hbar^2/GM^3).\label{vanmeter}\end{equation}
These results seem to suggest that the irreversible mechanism of such a collapse for the equation \eqref{NS} is of the same nature as the mechanism that leads to the appearance of stable solitons in the case of the nonlinear Schr\"odinger (NLS) equation \cite{CDW}: by radiating part of their mass (norm), wave packets diminish their energy until they reach a ground state (solitonic in the case of NLS) for which energy and norm conservation forbid further radiation.

\section{No-go theorems: Derrick's theorem.}
In his famous paper \cite{Derrick} Derrick sets out to show that stable traveling wave type solutions cannot exist in three or more dimensions. To this end he uses what is essentially a variational argument, but  as we shall see (and as was explained in great detail in \cite{CDW}), this argument should be treated with great care! As the mathematical reasoning in \cite{Derrick} is in fact applicable to equations in just a single space dimension (+time) as well, we choose a particular example on which it will become immediately clear not only that the entire reasoning is false, but also why that is the case.

Let us consider the one dimensional NLS equation 
\begin{equation}
i~\!\dfrac{\partial \psi}{\partial s} + \dfrac{\partial^2 \psi}{\partial z^2} + \big|\psi\big|^2 \psi = 0,\label{NLS}
\end{equation}
which  can be derived from a variational principle for the  action

\begin{equation}
\DIS A_{NLS}(\psi) = \iint_{-\infty}^{+\infty}~\!\left[\frac{{i}}{2}\big(\psi^* \frac{\partial \psi}{\partial s} - \psi \frac{\partial \psi^*}{\partial s}\big) - \big|\frac{\partial \psi}{\partial z}\big|^2 + \frac{1}{2} \big|\psi\big|^4\right] {d}z {d}s,\label{NLSac}
\end{equation}
and which is known to have remarkably stable localized (travelling) wave solutions: the so-called NLS solitons. It is a well known fact that (sufficiently fast decaying) initial conditions necessarily evolve towards trains of such solitons (while radiating excess energy) and that the solitons are not only stable for small perturbations but even scatter completely elastically among themselves.

The action \eqref{NLSac} is invariant under gauges $\psi(z,s)\mapsto e^{{i}\mathrm\varepsilon} \psi(z,s)$, translations $z\mapsto z+ \zeta$ and $s\mapsto s + \sigma$, and Galilean transformations: $(z,s,\psi) ~\mapsto~(z- v s,\, s,\, {e}^{-{i}(\frac{v}{2} z + \frac{v^2}{4} s)}\psi\,).$ By Noether's theorem, the first three of these symmetries (of the action) give rise to the following conserved quantities for NLS:
\begin{eqnarray*}
N(\psi) = \int_{-\infty}^{+\infty}\!\! \big|\psi\big|^2 {d}z, \quad~ P(\psi) = ~\!{i}\!\!\int_{-\infty}^{+\infty}\!\!\!\big(\psi^* \frac{\partial \psi}{\partial s} - \psi \frac{\partial \psi^*}{\partial s}\big) {d}z\\
\text{and}\quad E(\psi) = \!\int_{-\infty}^{+\infty}\!\!\!\Big(\!\big|\frac{\partial \psi}{\partial z}\big|^2 - \frac{1}{2} \big|\psi\big|^4\Big) {d}z.
\end{eqnarray*}
Moreover, the fourth variational symmetry (the Galilean invariance) yields the relation
$$\frac{{d} R}{{d}s} = \frac{P(\psi)}{N(\psi)}\qquad\text{for}\quad R(\psi) = \frac{1}{N(\psi)} \int_{-\infty}^{+\infty} z \big|\psi\big|^2 {d}z,$$
which shows that the center of mass of a solution to NLS moves with constant speed and hence, that any localized travelling wave solution $\big|\psi(z,s)\big|^2 = \big|\phi(z- v s)\big|^2$ can be put to rest by a Galilean transformation. It is therefore natural to look for stationary solutions of the form $\Psi(Z,S) = {e}^{{i}\beta S} \varphi(Z)$, which for some parameter $\beta\geq 0$ (if one wishes this solution to be localized in space) must satisfy:
$$\frac{\partial^2\varphi}{\partial Z^2} + \big|\varphi\big|^2 \varphi = \beta \varphi.$$
Derrick points out that the energy of such a stationary solution is (still) given by 
\begin{equation}
E(\varphi) = \int_{-\infty}^{+\infty}\left(\!\big|\frac{\partial \varphi}{\partial Z}\big|^2 - \frac{1}{2} \big|\varphi\big|^4\right) {d}Z, \label{NLSener}
\end{equation}
and then considers dilated functions $\DIS \varphi_\zeta(Z) = \varphi(\zeta Z)\,~\,(\zeta\in\mathbb{R}_{>0})$, with regard to which he wishes to minimize the energy. For this purpose he defines
$$ E_\zeta := \int_{-\infty}^{+\infty}~\left(\!\big|\frac{\partial \varphi_\zeta}{\partial Z}\big|^2 - \frac{1}{2} \big|\varphi_\zeta\big|^4\right) {d}Z ~= \zeta E_K + \frac{1}{\zeta} E_P,$$
in terms of a strictly positive `kinetic energy' term $\DIS E_K =\! \int_{-\infty}^{+\infty}\!\!\big|\frac{\partial \varphi}{\partial Z}\big|^2 {d}Z$ and a potential energy term $E_P = - \frac{1}{2} \int_{-\infty}^{+\infty}\!\!\big|\varphi\big|^4 {d}Z$ which is always strictly negative. In this  set-up, $\varphi(Z)=\varphi_1(Z)$ is supposedly `stable' with respect to dilations if (and only if) it minimizes $E_\zeta$ at $\zeta=1$. However, as the following simple calculation shows\footnote{One also finds that $\frac{d^2 E_\zeta}{d\zeta^2}\Big|_{\zeta=1} = 2 E_P \leq 0$, which would indicate that the energy can only reach a maximum at $\zeta=1$, instead of a minimum.}
$$ \frac{d E_\zeta}{d\zeta}\Big|_{\zeta=1} \equiv E_K - E_P = 0~\ \Leftrightarrow\ ~ E_K=E_P\,,$$
this is impossible as the kinetic and potential energies, by definition, have opposite signs.

The conclusion therefore seems to be that no stable traveling wave solution can exist for the NLS equation, which is obviously incorrect given the existence of soliton solutions for this equation. There is however a glaring oversight in Derrick's argument, which is that the equations he wants to discuss might have other variational symmetries (and conservation laws) than the two he needs in his variational set-up: time translational and Galilean invariance. Coming back to our example: although the NLS evolution is clearly ($L^2$) norm preserving, the dilated functions $\varphi_\zeta(Z)$ used in Derrick's variational argument do not preserve this norm. Hence, so-called `stability' with respect to such dilations is irrelevant when it comes to the real dynamics of the equation. Instead, one must formulate a variational argument for a class of functions with constant norm, which amounts to considering a variational principle with a Lagrange multiplier for the conserved norm.

In particular, the stationary NLS equation $\DIS \,\frac{\partial^2\varphi}{\partial Z^2} + \big|\varphi\big|^2 \varphi = \beta \varphi$ can be derived from the variational principle for the action
\begin{eqnarray*}
A = \int_{-\infty}^{+\infty} \left(\big|\frac{\partial\phi}{\partial Z}\big|^2 -\frac{1}{2} \big|\phi\big|^4\right){d}Z~+~\!\gamma \int_{-\infty}^{+\infty}\!\! \big|\phi\big|^2 {d}Z\\
\text{for}\quad\phi(Z)=\varphi(Z)\,,~\gamma=\beta.
\end{eqnarray*}
Consider now the class of functions $\DIS \varphi_\xi(Z) = \xi^{-1/2} \varphi(Z/\xi)$ and vary the action $A$ over $\varphi_\xi(Z)$, instead of the original dilations. This variation is now norm preserving. Requiring that $\varphi(Z) \equiv \varphi_1(Z)$ minimizes the action
$$A_\xi = \gamma N + \xi^{-2} E_K + \xi^{-1} E_P$$
at $\xi=1,$ and we obtain
$$ \frac{{d} A_\xi}{{d}\xi}\Big|_{\xi=1} = -(2 E_K + E_P) = 0~ \Leftrightarrow~\,E_P = -2 E_K\,,$$
for a kinetic and potential energy as defined above. Note that the stationary NLS equation tells us that $-E_K-2 E_P=\beta N$, and hence that 
$$E_K = \frac{\beta N}{3}\,,~\quad E_P = -\frac{2\beta N}{3}\qquad\Rightarrow\quad E = -\frac{\beta~\! N}{3}\leq0, $$
with energy $E$ as in \eqref{NLSener}.

Moreover, if we consider a rescaled function $\DIS \varphi_\eta(Z) = \eta^{-1} \varphi(Z/\eta)$, which can be shown by similar arguments to be still a solution of the stationary NLS equation but with $\eta$-dependent norm (and energy), we find that 
$$\beta= -\frac{E_K + 2 E_P}{N}\,\propto\, N^2,$$
so that the total energy scales as $N^3$ : $\DIS E = -\frac{\beta_1~\! N^3}{3}$ (for some $\beta_1\geq 0$).

In particular, the well-known solitonic solutions of NLS equation 
$$\varphi(Z) = \frac{\sqrt{2}\lambda}{\cosh(\lambda Z + \delta)}~\!,\qquad \beta=\lambda^2,$$
correspond to the values $N= 4~\! |\lambda|$,~$E = -\frac{4}{3}~\! |\lambda|^3$~~and $\beta_1 = \frac{1}{16}$.

Hence, the energy of a localized (single hump) solution to the stationary NLS (i.e., a stationary soliton) is indeed, a priori, unbounded: it decreases with increasing norm. The energy is however, de facto, bounded from below through the norm of the initial condition ($\sqrt{N_0}$  ):
$$E = -\frac{4}{3}~\! |\lambda|^3 \geq  -\frac{N_0^3}{48},$$
as the dispersive character of the NLS equation makes that any initial condition will lose energy and, necessarily norm, as it evolves in time (and, as a matter of fact, converges to a collection of solitons).
  
  A similar conclusion can be reached for many nonlinear evolution equations with analogous symmetry/conservation properties \cite{Kuznetsov,Turitsyn}: as soon as a nonlinear evolution equation possesses extra conservation laws (e.g. for the norm of the solution), Derrick's argument fails and a stable traveling wave becomes a possibility, in any dimension. For example, the Schr\"odinger-Newton  equation
 $$\displaystyle i \hbar\dfrac{\partial\Psi(t,{\bf x})}{\partial t}=-\dfrac{\hbar^2}{2m}\Delta\Psi(t,{\bf x})
-Gm^2\int_{\mathbb{R}^3} \dfrac{|\Psi(t,{\bf x'})|^2}{|{\bf x -x'}|}{d}^3 x'\, ~\Psi(t,{\bf x})$$
can be derived from the action
$$\DIS A_{SN}(\psi) = \iint\!{d}t {d}^3x~\!\Big[\frac{{i}\hbar}{2}\big(\psi^* (\bx,t)\frac{\partial \psi(\bx,t)}{\partial t} - \psi(\bx,t) \frac{\partial \psi^*(\bx,t)}{\partial t}\big)  $$
$$\DIS 
- \frac{\hbar^2}{2 M}\big|\nabla\psi(\bx,t)\big|^2
+ \frac{G M^2}{2} \int\!{d}^3y~\!\frac{\big|\psi(\by,t)\big|^2}{|\bx-\by|}~\!\big|\psi(\bx,t)\big|^2\Big] $$
as discussed in great detail in ref.\cite{CDW}. It obeys the same virial-like relation as the NLS equation for its energy : $2 E_K + E_P=0$, and has a `ground state' with total energy $E(\varphi) = -\frac{\mathrm{e}}{3} N(\varphi)^3  \frac{G^2 M^5}{\hbar^2}$ (for $e\approx 0.163$).

\section{Appendix: Born rule and no-signaling.}
Let us consider an alternative formulation of the standard quantum theory in which the probability of observing the outcome of an observable would not be equal to the square of the modulus of the projection of the initial quantum state, prior to measurement, onto the eigenspace associated to this outcome, but in which it would rather be proportional to another power of this modulus.\footnote{Born assumed e.g., in first instance, \cite{born} that the probability would be proportional to the modulus of the projection, not to its square, before he changed his mind when he expressed the rule carrying his name.} In this case, as we shall see, the no-signaling condition would no longer be satisfied. 
Technically, the proof is very simple. Let us consider, as in the preamble of Gisin's no-go theorem, a bipartite entangled system composed of a subsystem A and a subsystem B, while both subsystems are subject to local measurements.

For convenience we shall limit ourselves to a proof in which only pure states are involved, but the generalisation to non-pure states
(mixtures) is straightforward.  Let us assume that Alice and
Bob's systems are prepared in the pure (but not necessarily
factorizable) state $\ket{ \Psi}^{AB}= \sum_{i,j=0}
                ^{d-1}\alpha_{ij}\ket{ i}^A\otimes \ket{ j}^B$
(where $\ket{ i}^A$ and $\ket{ j}^B$ are states from
orthonormalized reference bases) and that Bob measures a local
observable in the $B$ region. Such an observable is represented by
 a local self-adjoint operator of the form $ Id.^A\otimes O^B$ so
  that its average value is equal to 
\begin{align*}
\sum_{i,j,i',j'=0}^{d-1}
  \alpha_{ij}^*\alpha_{i'j'}\bra{ i}^A\otimes\bra{ j}^B Id.^A&\otimes O^B
  \ket{ i'}^A\otimes
  \ket{ j'}^B \\  
  &=  \sum_{i,j,i',j'=0}^{d-1}\alpha_{ij}^*\alpha_{i'j'}\delta_{i,i'}\bra{ j}^B  O^B
  \ket{ j'}^B \\   
 &= \sum_{j,j'=0}^{d-1}\sum_{i=0}^{d-1}\alpha_{ij}^*\alpha_{ij'}\bra{ j}^B  O^B
  \ket{ j'}^B.
  \end{align*}

  It is worth noting that the results of Bob's local measurements are the same as those
   that he would obtain had he prepared his system in the state
  described by the effective or reduced density matrix\footnote{Actually, when the full state is entangled, there does not exist a local pure
 state that would reproduce the statistical distribution of local measurement outcomes.
  Instead, this statistics is described by the so-called reduced density matrix.
} 
  $\rho^B=\sum_{j,j'=0}^{d-1}\sum_{i=0}^{d-1}\alpha_{ij}^*\alpha_{ij'}
  \ket{ j'}^B\bra{ j}^B.$ Formally, this matrix can
   be obtained by tracing out external degrees of freedom
   (in this case Alice's degrees of freedom):

\begin{eqnarray*}
Tr_{A}(\ket{ \Psi}^{AB}\bra{ \Psi}^{AB}) & = & Tr_{A}(\sum_{i,j=0}
                ^{d-1}\alpha_{ij}\ket{ i}^A\otimes \ket{ j}^B\sum_{i',j'=0}
                ^{d-1}\alpha^*_{i'j'}\bra{ i'}^A\otimes \bra{
                j'}^B)\\
& = & \sum_{k=0}^{d-1}\bra{
k}^A(\sum_{i,j=0}^{d-1}\alpha_{ij}\ket{ i}^A\otimes \ket{
j}^B\sum_{i',j'=0}^{d-1}\alpha^*_{i'j'}\bra{ i'}^A\otimes \bra{
j'}^B)\ket{ k}^A \\ & = & \sum_{k=0}
                ^{d-1}(\sum_{i,j=0}
                ^{d-1}\alpha_{ij}\delta_{k,i} \ket{ j}^B\sum_{i',j'=0}
                ^{d-1}\alpha^*_{i'j'}\delta_{k,i'}\bra{ j'}^B \\
& = & \sum_{i,j,j'=0}^{d-1}\alpha_{ij} \ket{ j}^B
               \alpha^*_{ij'}\bra{ j'}^B=\rho^B.
\end{eqnarray*}

If Alice now imposes a local unitary transformation to her
qu$d$it, which sends the state $\ket{ i}^A$ onto $\ket{ \tilde
i}^A$, unitarity ensures that  $\braket{ i^A}{ i'^A}$ =
$\braket{\tilde i^A}{ \tilde i'^A}=\delta_{i,i'}$ so that, after
repeating the same (tilded) computation, we obtain the same
result. This shows that the average value of $O^B$ is not affected by
the unitary transformation imposed by Alice to the subsystem in
her possession.   Similarly, if Alice performs a measurement in
the $\ket{ i}^A$ basis, she will project Bob's state onto a state
proportional to the state
 $\sum_{j=0}^{d-1}\alpha_{ij}  \ket{ j}^B$
 with probability $P_{i}=\sum_{j=0}^{d-1}\vert \alpha_{ij}\vert^2$.
 The projector onto such a state, conveniently renormalized, is equal to
    ${ 1\over P_{i}}\sum_{j,j'=0}^{d-1}\alpha_{ij}  \ket{ j}^B\alpha^*_{ij'}
    \bra{ j'}^B$ so that after averaging over all possible outcomes
    of Alice ($i:0...d-1$), we obtain that $\langle O^B\rangle=
    \sum_{i=0}^{d-1}{ P_{i}\over P_{i}}\sum_{j,j'=0}^{d-1}\alpha_{ij}\alpha^*_{ij'}\bra{ j'}^B  O^B
  \ket{ j}^B$, which is equivalent to the average value that we derived in the absence of
  Alice's measurement.

In other words, Bob's reduced density matrix is independent of the measurement performed in A and/or of Alice's local modifications of the Hamiltonian $H_A$, and it is actually equal to the effective density matrix characterising the subsystem B in absence of any local measurement on the system A. This is the essence of the no-signaling condition. Now, it is easy to check that if we average the projectors on the collapsed states of the subsystem B, in accordance with a distribution $\tilde P_i$ which does not obey the Born rule, we find a reduced density matrix for the B system of the form $\langle O^B\rangle=
    \sum_{i=0}^{d-1}{ \tilde P_{i}\over P_{i}}\sum_{j,j'=0}^{d-1}\alpha_{ij}\alpha^*_{ij'}\bra{ j'}^B  O^B
  \ket{ j}^B$ which, generally, depends on whether or not one carries out a measurement on the A subsystem and also depends, in principle, on which local observable on the system A one chooses to measure. This means that when the Born rule is not respected, signaling occurs. In such circumstances (when the Born rule is not valid), in a frame where the collapse process is instantaneous, quantum correlations could in principle be used for sending classical information faster than light (which means supraluminal signaling or, as it is commonly called in the literature, signaling that constitutes ``spooky action-at-a-distance'', as originally considered by Einstein).

\vskip 30pt
\bibliographystyle{plain}

\end{document}